\documentclass[floatfix,aps,twocolumn,showpacs,nofootinbib,superscriptaddress]{revtex4}

\usepackage{graphicx}
\usepackage{epsfig}

\usepackage{amsmath}
\usepackage{amsfonts}
\usepackage{amssymb}
\usepackage{bm}
\usepackage{mathtools}

\usepackage{hyperref}
\usepackage{dsfont}
\usepackage{lipsum}

\usepackage{pdfpages}

\usepackage{tikz}
\usepackage{circuitikz}
\usetikzlibrary{arrows, shapes}
\usepackage{pgfplots}
\pgfplotsset{compat=1.14}

\newcommand{\figpath}[1]{./figures/#1}
\newcommand{\changescolor}{black}

\newcommand{\mean}[1]{\left\langle #1 \right\rangle}
\newcommand{\meann}[1]{\langle #1 \rangle}

\begin{document}

\author{Nahuel Freitas}
\affiliation{Complex Systems and Statistical Mechanics, Department of Physics and Materials Science,
University of Luxembourg, L-1511 Luxembourg, Luxembourg}
\author{Jean-Charles Delvenne}
\affiliation{Institute of Information and Communication Technologies, Electronics and
Applied Mathematics, Universit\'e catholique de Louvain, Louvain-La-Neuve, Belgium}
\author{Massimiliano Esposito}
\affiliation{Complex Systems and Statistical Mechanics, Department of Physics and Materials Science,
University of Luxembourg, L-1511 Luxembourg, Luxembourg}

\title{Stochastic Thermodynamics of Non-Linear Electronic Circuits: \\
A Realistic Framework for Computing around kT}

\date{\today}

\begin{abstract}
We present a general formalism for the construction of thermodynamically
consistent stochastic models of non-linear electronic circuits.
The devices constituting the circuit can have arbitrary I-V curves and may include
tunnel junctions, diodes, and MOS transistors in subthreshold operation, among others.
We provide a full analysis of the stochastic non-equilibrium thermodynamics of these models,
identifying the relevant thermodynamic potentials, characterizing the different contributions
to the irreversible entropy production, and obtaining different fluctuation theorems.
Our work provides a realistic framework to study thermodynamics of computing with electronic circuits.
We demonstrate this point by constructing a stochastic model of a CMOS inverter.
We find that a deterministic analysis is only compatible with the assumption of equilibrium fluctuations,
and analyze how the non-equilibrium fluctuations induce deviations from its
deterministic transfer function. Finally, building on the CMOS inverter, we propose
a full-CMOS design for a probabilistic bit (or binary stochastic neuron) exploiting
intrinsic noise.
\end{abstract}

\maketitle

\section{Introduction}
\label{sec:introduction}

The growing energy consumption of data-intensive technologies is raising
concern \cite{mora2018, masanet2020}.
Semiconductor-based electronic circuits are the dominant technology for information processing.
In order to reduce energy consumption, those circuits
need to function in regimes where the information carrying signals are increasingly close to
thermal fluctuations \cite{kish2002, sadek2003, krishnan2007, kish2009, hamilton2014, orlov2019, rezaei2020}.
Indeed, proposals have been made to trade reliability for energetic costs in certain applications
\cite{palem2005, cheemalavagu2005, han2013}, or to exploit intrinsic thermal fluctuations in order
to solve stochastic optimization problems \cite{camsari2017, borders2019, kaiser2020}.
But while non-linear electronic circuits are essential for computation, the proper description
of thermal noise in these circuits is a highly non-trivial problem, especially if one aims at
preserving thermodynamic consistency \cite{blanter2000, mcfee1971, vankampen1960, weiss1998, wyatt1997, wyatt1999, brillouin1950}.
Traditional methods employed in engineering for the description of noise are usually based on the
linearization of a given element response around an operating point, or only consider external noise
generated by linear resistors (the internal or intrinsic noise is first mapped to external sources)
\cite{weiss1998, tan1985}. Although these approaches might offer accurate estimations of
the noise in some applications, they are not thermodynamically consistent and therefore are not
suited for situations in which thermal fluctuations are relevant or, in particular,
are exploited as a resource.

The discovery of the so-called fluctuation theorems
\cite{esposito2009, campisi2011, jarzynski2011, seifert2012, rao2018detailed}
and the ensuing development of the theory of stochastic thermodynamics
\cite{sekimoto2010, seifert2012, zhang2012, rao2018conservation},
established very general constraints on the thermal fluctuations of different physical systems,
even if they are highly non-linear and arbitrarily away from thermal equilibrium.
The theory was used to study colloidal particles, chemical reaction networks, molecular motors
\cite{ciliberto2017, rao2018, parrondo2002, seifert2012}, but also in electronic systems.
There, it was applied to linear electrical circuits ranging from simple circuits
\cite{van2004, garnier2005, ciliberto2013} to complex networks (even in quantum regimes) \cite{freitas2019}, but these circuits are of limited use to implement computations.
Also, after experimental and theoretical progress on the study of non-linear single electron devices and
Coulomb blockade
systems \cite{devoret1990, blanter2000, wasshuber2012, bagrets2003, averin1986, ali1994},
stochastic thermodynamics was successfully applied to reach a deep understanding of thermal fluctuations
in these systems \cite{andrieux2006mesoscopic, esposito2007, cuetara2011, koski2013, cuetara2015, pekola2015, benenti2017, gu2019}.
In those devices, the nature of the conduction channels (typically tunnel junctions)
and the nanoscopic size of the conductors (which implies low capacitances and high charging energies),
allow to design circuits that process information with very low energy requirements \cite{wasshuber2012}.
The logical states are represented here by the presence of just one or few electrons.
Unfortunately, the requirement of low temperatures and challenges in the fabrication of
these devices have so far prevented their practical application in computing.
In the foreseeable future, regular complementary metal-oxide-semiconductor (CMOS)
circuits will probably remain the most relevant platform for computing \cite{enz2006, tsividis2011}.
In this context, the quest for speed and low energy consumption fueled a spectacular progress in the
miniaturization of transistors, and nowadays integrated circuits with typical features size of
around $5\text{nm}$ can be mass produced.
In these circuits, a transistor can be activated by just a few hundred electrons in its gate terminal.
Thus, CMOS circuits are approaching regimes where a description in terms of single
electrons becomes relevant \cite{rezaei2020}.
The fact that both single electron devices and CMOS transistors (in some modes of
operation) display shot noise \cite{sarpeshkar1993}, suggests that the rigorous tools and methods that have been used for
the modeling and simulation of single electron devices could also be applied to study CMOS circuits.
While recent studies used stochastic thermodynamics for the detailed characterization
of individual devices such as diodes and transistors \cite{gu2018, gu2019, gu2020},
the study of complex networks and circuits comprising such devices remains unexplored.

Stochastic thermodynamics also offers a framework to study the
energetic costs associated to computing in a systematic way \cite{wolpert2019, sagawa2019}.
Although it was realized early on that information processing can in
principle be done without energy expenditure \cite{bennett1982, bennett1985},
this is only true in idealized setups in which either the computation is extremely slow
or the computing device is perfectly isolated from the environment.
Practical (artificial or natural) computing done in finite time and in noisy environments
involves the dissipation of energy. In this context, the modern theory of stochastic
thermodynamics was employed to clarify early notions like the relation between
thermodynamic and logical reversibility \cite{sagawa2014, sagawa2019, wolpert2019},
the energetic costs associated to measurement
and erasing \cite{sagawa2009, sagawa2014, horowitz2014, ptaszynski2019},
\textcolor{\changescolor}{and the cyclic operation of computing devices \cite{wolpert2019}}.
Also, stochastic thermodynamics
allows to rigorously evaluate the dissipation of nonequilibrium
and finite-time processes \cite{aurell2012, schmiedl2007, sivak2012, nicholson2020},
which has applications for optimal erasing protocols
\cite{diana2013, zulkowski2014, proesmans2020, sheng2019}.
The thermodynamic costs associated to the structure of complex information processing networks were
also studied \cite{boyd2018, wolpert2020}. However, these efforts involve either
extremely simple and idealized models or abstract formulations aimed at obtaining fundamental
bounds, where no particular computing architecture is considered.

In this paper we report three major achievements.
First, in sections II to IV, we resolve the long standing and fundamental problem of
rigorously describing noise in non-linear electronic circuits.
We do so by developing a general formalism to construct \emph{thermodynamically consistent}
stochastic models of \emph{arbitrary} non-linear circuits.
Despite the presence of charging effects causing the behavior of each
device to be modified by its environment \cite{devoret1990}, given the I-V curve
characterization of a single arbitrary device,
we show how to describe its stochastic behavior when it is introduced
in an arbitrary circuit, avoiding some pitfalls leading to unphysical steady states.
In this way we can seamlessly accommodate many different devices like tunnel junctions
\cite{wasshuber2012,koski2013}, diodes \cite{wyatt1997}, MOS transistors in subthreshold
operation \cite{wyatt1997, sarpeshkar1993, wang2006}, or more exotic devices like nanoscale vacuum channel
transistors \cite{han2012}. The devices may even be time-dependently driven and at different temperatures.

Second, in sections V and VI,
we provide a complete thermodynamic characterization of these non-linear electronic
circuits using stochastic thermodynamics.
We formulate the first and second law of thermodynamics for these circuits at the ensemble
averaged as well as at the fluctuating level.
In doing so we establish the relation between heat and electric currents
and identify the thermodynamic potentials and forces at work in these circuits.
We also formulate a general version of the Landauer principle as well as
the different fluctuation theorems known to date.

Third, by formulating a stochastic thermodynamics describing a large family
of technological relevant electronic circuits, we provide a rigorous framework to study
thermodynamics of computation implemented with realistic architectures instead of toy models.
To assert this claim, in section VII we construct and analyze a stochastic model of a
\emph{CMOS inverter} (or NOT gate) and of a \emph{probabilistic bit} ($p$-bit).
The CMOS inverter is an important primitive in electronic design,
from which more complex devices like oscillators and memories can be built.
We show how, due to non-equilibrium fluctuations, the transfer function
of the gate deviates from the one obtained by a deterministic treatment.
We also compute the full counting statistics of the current fluctuations
and illustrate the validity of a detailed fluctuation theorem.
The $p$-bit can be considered as a faulty memory, with a controllable bias and error rate.
They are a physical implementation of what in machine
learning is known as a \emph{binary stochastic neuron} \cite{ackley1985, bengio2013}.
Such devices were recently employed in proof-of-concept experiments to solve
stochastic optimization problems and emulate artificial neural networks \cite{camsari2017, borders2019}.
Other proposals to exploit physical noise include cryptographic applications \cite{nawaz2019}.
To the best of our knowledge, our design is the first full-CMOS proposal for a $p$-bit
exploiting intrinsic noise, and can be implemented with current technology.
The methods employed in these examples can be directly applied to model
arbitrary logic gates at the stochastic level\textcolor{\changescolor}{, both in asynchronous
computing schemes or synchronous ones requiring an external clock signal.}

Our work provides bridges between computer engineering, mesoscopic physics and nonequilibrium
statistical physics. In doing so it may contribute to the search for new practical
and energy-efficient computing paradigms, and also to the design of experiments taking
advantage of the versatility of electronic circuits in order to test new developments
in statistical physics.

Please note that the reader mainly interested in the stochastic modeling of
circuits might want to skip Sections V-VI on a first read, and go directly to the applications
discussed in Section VII.

\section{Basic Setup}

\begin{figure}
\centering
\includegraphics[scale=.20]{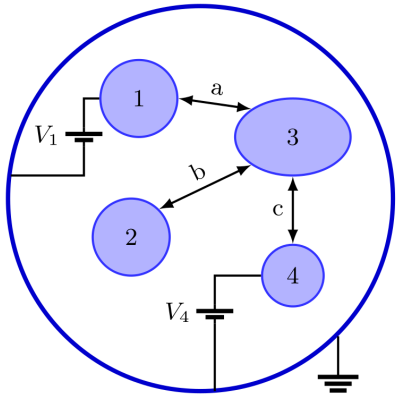}
%
%
\caption{A system of conductors. Two of them, 1 and 4, are maintained at fixed potentials
$V_1$ and $V_4$ by voltage sources. Elementary charges $q_e$ are interchanged between them by devices
$a$, $b$ and $c$.}
\label{fig:conductors}
\end{figure}

We consider an arrangement of $N_0$ ideal conductors characterized by their total charge
$\{q_n\}_{n=1,\cdots,N_0}$ and electrostatic potential $\{V_n\}_{n=1,\cdots,N_0}$, where the
latter are measured with respect to some reference or `ground' (see Figure \ref{fig:conductors}).
Basic electrostatic theory shows that the charges and potentials are related by the linear relation:
\begin{equation}
\bm{q_0} = \bm{C_0} \bm{V_0},
\label{eq:capacitance_matrix}
\end{equation}
where  $\bm{q_0} = (q_1, \cdots, q_{N_0})^T$ and
$\bm{V_0}= (V_1, \cdots, V_{N_0})^T$ are column vectors
containing the charges and voltages, and the $N_0\times N_0$ symmetric matrix $\bm{C_0}$
(known as the Maxwell capacitance matrix) encodes the mutual and self capacitances
of the conductors.
These capacitances depend on the shape, size, and relative position and
orientation of the conductors. The electrostatic energy contained in such a system
is given by the quadratic form:
\begin{equation}
	E = \frac{1}{2} \bm{q_0}^T \bm{V_0} = \frac{1}{2} \bm{V_0}^T \bm{C_0} \bm{V_0}
    = \frac{1}{2} \bm{q_0}^T \bm{C_0}^{-1} \bm{q_0}.
\end{equation}
Some of the conductors can have their potential fixed by voltage sources, and in
that case we say that the circuit is \emph{open}.
We will refer to conductors with fixed potentials as \emph{regulated}
conductors, and to the rest as \emph{free} conductors.
Thus, we have $N_0=N+N_r$, where $N$ is the number of free conductors and $N_r$
the number of regulated conductors. The vector $\bm{q_0}$ will be split into vectors
$\bm{q}$ and $\bm{q}_r$ containing the charges of the free and regulated conductors,
respectively.
If the circuit is open, then the degrees of freedom of the system are reduced. This can be seen
from Eq. \eqref{eq:capacitance_matrix}, since fixing the potential of a conductor
imposes a linear relationship between all the charges. Then, the state
of the system is fully specified by the charges $\bm{q}$ of all the free conductors.

We also consider two-terminal devices or channels that allow the transport
of elementary charges of value $q_e$ between pairs of conductors, forming
a network or circuit.
Each of these channels is modeled as a bi-directional Poisson process (BPP).
This choice offers some generality
while keeping the formalism concrete and simple and, as mentioned before, allows to
describe relevant devices like tunnel junctions \cite{wasshuber2012,koski2013},
diodes \cite{wyatt1997}, MOS transistors in subthreshold
operation \cite{sarpeshkar1993, wang2006}, and other devices like nanoscale vacuum channel
transistors \cite{han2012}. The common feature of all these devices is that
they display shot-noise (i.e., the current noise spectral density is proportional
to the average current for large biases \cite{scholten2006}, see section \ref{sec:IV_curve}).
This is also a limitation of the BPP modeling choice, since it does not
allow to properly describe regular resistors or source-drain conduction
in MOS transistors operating in saturation mode, where the current noise spectral
density is approximately independent of the average current (as for Johnson-Nyquist noise)
\cite{scholten2006}. However this is not a serious limitation,
especially if we are interested in the regime of ultralow energy consumption, where the
subthreshold and unsaturated operation of MOS transistors is optimal \cite{wang2006}.

Let $\rho=1,\cdots,M$ index the two-terminal devices present in the circuit.
Then, given a device $\rho$ connecting conductor $n$ and $m$, we associate to it
two basic Poisson processes: a `forward' one in which an elementary charge is
transported from conductor $n$ to conductor $m$, and the `reverse' one in which a
charge is transported in the opposite direction, with respective rates
$\lambda_\rho(\bm{q},t)$ and $\lambda_{-\rho}(\bm{q},t)$. The forward direction is of course
chosen arbitrarily. Note that the rates $\lambda_{\pm \rho}(\bm{q},t)$ depend explicitly
on the full state $\bm{q}$ of the system, which allows to model externally controlled
conduction channels.
Thus, in a transition
$\pm \rho$  the state of the system changes as:
\begin{equation}
\bm{q} \to \bm{q} + q_e \bm{\Delta}_{\pm\rho} \qquad \qquad (\bm{\Delta}_\rho)_k =
-\delta_{k,n} + \delta_{k,m},
\end{equation}
where $q_e$ is the value of the elementary charge involved in all the possible
transitions and $\bm{\Delta}_{-\rho} = -\bm{\Delta}_{\rho}$ .
The vector $\bm{\Delta}_\rho$ encodes to which conductors the
device $\rho$ is connected and what is the change in their number of charges
during the transitions.
If a device $\rho$ is connected between one of the free conductors, $n$, and one with fixed voltage,
the corresponding vector $\bm{\Delta}_\rho$ is given by:
\begin{equation}
(\bm{\Delta}_\rho)_k =  \delta_{k,n},
\end{equation}
where the forward direction was chosen as the one leaving the conductor with
fixed potential.

One can imagine more complex devices that involve three
or more conductors in an irreducible way, for example by taking one charge from
conductor $m$, one from $n$, and transporting them to conductor $o$. This kind
of devices can also be treated with the formalism we develop here, although
most discussions will be focused on two-terminal devices (that however might be controlled
externally).

\subsection{Reduced incidence matrix, cycles and conservation laws}
\label{sec:cycles_and_cons_laws}

The vectors $\bm{\Delta}_{\rho>0}$ can be grouped in an $N\times M$ \emph{reduced incidence matrix}:
\begin{equation}
\bm{\Delta} =
\left[
\begin{matrix}
\vert & \cdots & \vert \\
	\bm{\Delta}_1 & \cdots & \bm{\Delta}_M \\
\vert & \cdots & \vert \\
\end{matrix}
\right].
\end{equation}
This matrix is analogous to the stoichiometric matrix in chemical reaction
networks \cite{polettini2014, rao2018}\footnote{A given circuit with
two-terminal devices can be mapped to a chemical reaction network by mapping
conductors to chemical species $A, B, C,\cdots $ and devices to chemical reactions $X \leftrightarrow Y$. This mapping only concerns the structure of the circuit and network, not
the dynamics or thermodynamics.}.
For closed circuits it coincides with the full incidence matrix of the directed graph
obtained by mapping conductors to nodes and two-terminal devices as directed edges
(with the direction given by the forward one). The reduced incidence matrix for an open
circuit is obtained from the one of the closed circuit by eliminating the rows
corresponding to regulated conductors.

The right null eigenvectors of $\bm{\Delta}$ define \emph{cycles}, i.e., sequences of
transitions that leave the circuit state invariant:
\begin{equation}
\bm{\Delta} \: \bm{c}_\alpha = 0.
\end{equation}
The elements of the vectors $\bm{c}_\alpha$ can always be chosen to be $0$,$1$, or $-1$.
The number of independent cycles is ${N_c = \text{dim}(\text{Ker}(\bm{\Delta}))}$.
The left null eigenvectors of $\bm{\Delta}$ correspond to conservation laws, since if
\begin{equation}
\bm{\ell}_\nu^T \bm{\Delta} = 0,
\end{equation}
then the quantities
\begin{equation}
L_{\nu}(\bm{q}) = \bm{\ell}_{\nu}^T \bm{q}
\label{eq:conserved_quantities}
\end{equation}
will not change under any transition, i.e., they are determined by the initial
state of the circuit. The elements of $\bm{\ell}_\nu$ can always be considered
to be $0$ or $1$. For each connected component of the full circuit in
which no conductor is regulated, we have a conserved quantity
that is just the total charge of the conductors in that component. In fact,
these are the only conserved quantities. Thus, the number of independent
conservation laws, $N_l = \text{dim}(\text{Ker}(\bm{\Delta}^T))$, equals the number
of closed connected components of the circuit.

Whenever a closed circuit is opened by connecting one of its conductors to
a voltage source (see Figure \ref{fig:conductors}),
we might either break a conservation law or create a new cycle.
This can be seen in the following way. The rank-nullity theorem applied to
the matrix $\bm{\Delta}$ can be expressed as:
\begin{equation}
N - N_l + N_c = M.
\end{equation}
This is valid for closed as well as for open circuits. Let us assume however that
in the previous equation $N$, $N_l$ and $N_c$ correspond to the matrix $\bm{\Delta}$
of the circuit in which all the voltage sources are disconnected.
Then we connect $N_r$ voltage sources, and thus the number of conductors involved
in the new matrix $\bm{\Delta}'$ is now $N'=N-N_r$. Applying the rank-nullity theorem
to $\bm{\Delta}'$ we obtain:
\begin{equation}
N' - N'_l + N'_c = M.
\end{equation}
Subtracting the previous two equations we see that:
\begin{equation}
N_r = N_l - N'_l + N'_c - N_c.
\end{equation}
Thus, the number of voltage sources connected to the circuit equals the
number of broken conservation laws, $N_l - N_l'$, plus the number of
emergent cycles, $N'_c - N_c$. This is easily understood: if a previously
closed component of the circuit is connected to a voltage source, its
total charge ceases to be a conserved quantity. However, if we further
connect another voltage source to another conductor of the same component,
then a new cycle is created (the one in which a charge is injected by
one source, transported through the component, and removed by the second source).

\subsection{Stochastic and deterministic dynamics}

At any given time the state of the circuit is described by a probability
distribution $P(\bm{q},t)$ over the state space. It evolves according to the
master equation:
\begin{equation}
d_t P(\bm{q},t) = \sum_{\rho} \left\{ J_{-\rho}(\bm{q}+q_e\bm{\Delta}_\rho, t) - J_{\rho}(\bm{q},t) \right\},
\label{eq:master_eq}
\end{equation}
where the probability currents are defined as:
\begin{equation}
J_\rho(\bm{q},t) = \lambda_\rho(\bm{q},t) P(\bm{q},t).
\label{eq:prob_currents}
\end{equation}
They are simply the probability per unit time to observe a transition $\rho$
in state $\bm{q}$. The summation in Eq. \eqref{eq:master_eq} is over positive and
negative values of $\rho$, i.e., it is over transitions and not over devices.
The set of currents ${J_{\rho}(\bm{q},t)}_{\rho=\pm1,\cdots,\pm M}$ can be considered
the components of a vector function of the state, that we denote $\bm{\mathcal{J}}(\bm{q},t)$.
We define the following operator over those functions:
\begin{equation}
D_{\bm{q}}^\rho[\bm{\mathcal{F}}] = F_{-\rho}(\bm{q}+q_e\bm{\Delta}_\rho) - F_{\rho}(\bm{q}).
\label{eq:def_D}
\end{equation}
Then, $D_{\bm{q}}^\rho[\bm{\mathcal{J}}]$ is the net probability current arriving at state $\bm{q}$
corresponding to transitions $\pm \rho$, and the master equation reads:
\begin{equation}
d_t P(\bm{q},t) = \sum_{\rho} D^\rho_{\bm{q}}[\bm{\mathcal{J}}].
\label{eq:master_eq_D}
\end{equation}
Note that the change in a scalar quantity $F(\bm{q})$ in a transition
$\bm{q} \to \bm{q} + q_e \bm{\Delta}_\rho$ can also be expressed by trivially
extending the operator $D_{\bm{q}}^\rho [\cdot]$ to these functions:
\begin{equation}
D_{\bm{q}}^{\rho}[F] = F(\bm{q}+q_e \bm{\Delta}_\rho) - F(\bm{q}).
\end{equation}

The dynamical description based on the master equation in Eq. \eqref{eq:master_eq}
is valid for timescales which are large compared to the time taken by each transition
or conduction event, that here are considered to be instantaneous.
This dynamics can be compared to the
deterministic dynamics obtained by usual methods in circuit theory \cite{chua1987}.
In those deterministic descriptions, the charges $q_n$ are considered to be continuous
variables, and the charge vector $\bm{q}$ evolves according to the following
differential equation:
\begin{equation}
d_t \bm{q} = \sum_{\rho>0} \bm{\Delta}_\rho \: I_\rho,
\label{eq:deterministic}
\end{equation}
where $I_\rho$ is the electric current associated to device $\rho$. The previous
equation is closed by providing the I-V curve characterization of all the devices,
and by Eq. \eqref{eq:capacitance_matrix} relating voltages and charges. For example,
the current $I_\rho$ through a two-terminal device connected from conductor $n$ to conductor $m$
is considered to be a function $I_\rho(\Delta V_{n,m})$ of the
voltage drop $\Delta V_{n,m} = V_n - V_m$ (see Section \ref{sec:IV_curve}).
Then, $\Delta V_{n,m}$ can be expressed in terms of $\bm{q}$ by
inverting Eq. \eqref{eq:capacitance_matrix}.
The relation between the deterministic and stochastic descriptions is non-trivial
and will be examined in the particular example of the CMOS inverter in section
\ref{sec:cmos_inverter}.

\subsection{Equilibrium states and detailed balance}

An equilibrium state $P_\text{eq}(q,t)$ of the circuit is defined as one in which
the global detailed balance condition holds:
\begin{equation}
D^\rho_{\bm{q}}[\bm{\mathcal{J}}] = 0 \qquad \forall \rho.
\label{eq:detailed_balance}
\end{equation}
By Eq. \eqref{eq:master_eq_D}, if an equilibrium state exists it is also a stationary one.
In general no equilibrium state exists. However, for closed and isothermal
circuits consistency with equilibrium thermodynamics demands the following
Gibbs state to be an equilibrium one:
\begin{equation}
P_\text{Gibbs}(\bm{q},t) = Z^{-1}\: e^{-\beta E(\bm{q})}\: \prod_\nu \delta [L_\nu(\bm{q}), L_\nu(\bm{q}^{(i)})],
\label{eq:gibbs_state}
\end{equation}
where $\delta[a,b] = 1$ if $a=b$ and $0$ otherwise, $\bm{q}^{(i)}$ is the initial state
of the circuit, and $\nu$ runs over a set of independent conservation laws.
The partition function $Z$ is such that $P_\text{Gibbs}(\bm{q},t)$ is normalized
and thus depends on the inverse temperature $\beta$ and the quantities
$\{L_\nu(\bm{q}^{(i)})\}$. More general equilibrium states can be obtained by
mixing Gibbs distributions like Eq. \eqref{eq:gibbs_state} according to a distribution
$P(\bm{q}^{(i)})$ on the initial state.

The demand that $P_\text{Gibbs}(\bm{q},t)$ must be an equilibrium
state when the circuit is closed and isothermal imposes minimal conditions
on the transition rates $\lambda_{\pm \rho}(\bm{q},t)$. These are the \emph{local}
detailed balance (LDB) conditions which for closed circuits and isothermal settings are:
\begin{equation}
\log \frac{\lambda_\rho(\bm{q},t)}
{\lambda_{-\rho}(\bm{q}+q_e\bm{\Delta}_\rho,t)}
= -\beta \left( E(\bm{q}+q_e\bm{\Delta}_\rho) - E(\bm{q})\right),\\
\label{eq:ldb_closed_iso}
\end{equation}
for each $\rho$.
They can also be written as:
\begin{equation}
\begin{split}
D_{\bm{q}}^{\rho}[\log \bm{\lambda}] &= \beta D_{\bm{q}}^{\rho}[E],
\end{split}
\end{equation}
where $\bm{\lambda}(\bm{q},t)$ is a vector function of the state with components
$\{\lambda_\rho(\bm{q},t)\}_{\rho=\pm 1,\cdots,\pm N}$, and
the $\log(\cdot)$ function is applied element-wise. Thus,
the rates $\lambda_{\pm\rho}(\bm{q},t)$ characterizing
a given two-terminal device $\rho$
must fulfill the constraints imposed by Eq. \eqref{eq:ldb_closed_iso}.
We now generalize the LDB conditions to open circuits
and non-isothermal settings.

\subsection{Energy difference and local detailed balance}
\label{sec:energy_diff}

We consider an open circuit in which some conductors have the potential fixed
by voltage sources.
In the same way we did with the charges, the vector $\bm{V_0}$ is split into vectors $\bm{V}$ and $\bm{V}_r$, containing the voltages of the free and regulated conductors, respectively.
We can then express the relation of
Eq. \eqref{eq:capacitance_matrix} between all the charges and voltages as:
\begin{equation}
\left[
\begin{matrix}
	\bm{q}\\
	\bm{q}_r
\end{matrix}
\right]
=
\left[
\begin{matrix}
	\bm{C} & \bm{C}_m\\
	\bm{C}_m^T & \bm{C}_r
\end{matrix}
\right]
\left[
\begin{matrix}
	\bm{V}\\
	\bm{V}_r
\end{matrix}
\right],
\label{eq:block_C_matrix}
\end{equation}
where $\bm{C}_r$ is the $N_r\times N_r$ capacitance matrix of the
regulated conductors, $\bm{C}$ is the $N\times N$ capacitance matrix
of the free conductors,
and $\bm{C}_m$ is the $N\times N_r$ matrix with the mutual capacitances
between conductors of the two groups. The previous equation can be rewritten as:
\begin{equation}
  \left[
\begin{matrix}
	\bm{V}\\
	\bm{q}_r
\end{matrix}
\right]
=
\left[
\begin{matrix}
	\bm{C}^{-1} & & -\bm{C}^{-1}\bm{C}_m\\
	\bm{C}_m^T\bm{C}^{-1} & & \bm{C}_r - \bm{C}_m^T \bm{C}^{-1} \bm{C}_m
\end{matrix}
\right]
\left[
\begin{matrix}
	\bm{q}\\
	\bm{V}_r
\end{matrix}
\right],
\label{eq:C_matrix_split}
\end{equation}
from where it is clear that, given the potentials $\bm{V}_r$, the charges $\bm{q}$ are
enough to determine the rest of the variables, as discussed before. The
total electrostatic energy is then:
\begin{equation}
\begin{split}
	E(\bm{q}) &= \frac{1}{2}
\left[
\begin{matrix}
	\bm{q}^T, \bm{V}_r^T
\end{matrix}
\right]
\left[
\begin{matrix}
	\bm{V}\\
	\bm{q}_r
\end{matrix}
\right] \\
	&= \frac{1}{2} \bm{q}^T \bm{C}^{-1} \bm{q} + \frac{1}{2} \bm{V}_r^T (\bm{C}_r - \bm{C}_m^T\bm{C}^{-1}\bm{C}_m) \bm{V}_r.
\end{split}
\label{eq:internal_energy}
\end{equation}
We are interested in computing how the total energy of the system (conductors plus
sources) changes in a transition $\bm{q} \to \bm{q}+q_e\bm{\Delta}_\rho$. From the previous equation we see that
the change in electrostatic energy is
\begin{equation}
\begin{split}
	D^{\rho}_{\bm{q}}[E] &=
	E(\bm{q}+q_e \bm{\Delta}_\rho) - E(\bm{q}) \\
	&= q_e^2\: \bm{\Delta}_\rho^T \bm{C}^{-1} \bm{\Delta}_\rho/2
	+ q_e \:\bm{q}^T \bm{C}^{-1} \bm{\Delta}_\rho,
\end{split}
\end{equation}
which is independent of the voltages $\bm{V}_r$. In addition to this,
we need to consider the change in the energy stored in the voltage sources.
This can be be computed as (minus) the work performed by them, which equals
the charge transported from ground to the conductor to which each source is
connected times its voltage. There are two different contributions to this work.
First, the transition $\rho$ might directly involve one regulated conductor.
If a charge $q_e$ arrives to the conductor fixed to a potential $V_r$, it needs
to be removed, and for this the source must perform an amount of work given by
$w_r = -q_e V_r$. Secondly, even if the transition does not involve any regulated
conductor, changes in the distribution of charge among the free conductors
can induce a charging of the regulated conductors. From Eq. \eqref{eq:C_matrix_split}
we see that the induced charge is $\delta \bm{q}_r = q_e \bm{C}_m^T \bm{C}^{-1} \bm{\Delta}_\rho$.
It follows that the total amount of work performed by the sources during
transition $\rho$ is
\begin{equation}
	\delta W_\rho =  - q_e\bm{V}_r^T\bm{\Delta}_\rho^r + q_e \bm{V}_r^T\bm{C}_m^T \bm{C}^{-1}\bm{\Delta}_\rho,
\end{equation}
where $\bm{\Delta}_\rho^r$ is a vector encoding the change in the number of charges
in the regulated conductors in transition $\rho$ (if no regulated conductor is
involved in transition $\rho$, then $\bm{\Delta}_\rho^r = 0$).
Thus, the change in the energy of the system and sources can be written as:
\begin{equation}
\begin{split}
\delta Q_\rho(\bm{q}) &= D^{\rho}_{\bm{q}}[E] - \delta W_\rho = D^{\rho}_{\bm{q}}[\Phi] + q_e \bm{V}_r^T\bm{\Delta}_\rho^r,\\
\end{split}
\label{eq:delta_Q}
\end{equation}
where we have defined the potential
\begin{equation}
	\Phi(\bm{q}) =  E(\bm{q})-\bm{V}_r^T\bm{C}_m^T\bm{C}^{-1}\bm{q}.
  \label{eq:phi}
\end{equation}
The first term in the right hand side of Eq. \eqref{eq:delta_Q} is conservative,
since its contribution vanishes
in any cyclic sequence of transitions in the state space $\{\bm{q}\}$. The second
contribution is not conservative since its contribution does not vanish in cyclic transformations:
its value will depend on how the regulated conductors are involved in the cycle.
Also, we note that the gradient of the potential $\Phi(\bm{q})$ gives the voltage of
the free conductors:
\begin{equation}
\bm{V}(\bm{q}) = \nabla_{\bm{q}} \Phi(\bm{q}),
\label{eq:gradient_phi}
\end{equation}
as can be verified from Eq. \eqref{eq:C_matrix_split}.

The quantity $\delta Q_{\rho}(\bm{q})$ is the energy required to perform the transition
$\bm{q}\to \bm{q}+q_e\bm{\Delta}_\rho$. By conservation of energy, it must be provided by the environment of the device $\rho$, which
we assume to be at thermal equilibrium at temperature $T_\rho$ (this
implies that the two conductors to which $\rho$ is connected should also be at
temperature $T_\rho$). Therefore, $\delta Q_{\rho}(\bm{q})$ is the heat associated
to device $\rho$ during the transition, and corresponds to an entropy change in its
environment equal to $-\delta Q_{\rho}(\bm{q})/T_\rho$. Thus, the LDB condition now reads:
\begin{equation}
\begin{split}
	\log \frac{\lambda_\rho(\bm{q}, t)}{\lambda_{-\rho}(\bm{q}+
  q_e\bm{\Delta}_\rho,t)}
	&= -\beta_\rho \: \delta Q_\rho(\bm{q}),
\label{eq:ldb}
\end{split}
\end{equation}
for each $\rho$, where $\beta_\rho = (k_b T_\rho)^{-1}$. Equivalently,
\begin{equation}
D_{\bm{q}}^{\rho}[\log \bm{\lambda}]
= \beta_\rho \left( D_{\bm{q}}^{\rho}[\Phi] + q_e \bm{V}_r^T \bm{\Delta}_\rho^r\right).
\label{eq:ldb_compact}
\end{equation}
For closed and isothermal settings this condition reduces to Eq. \eqref{eq:ldb_closed_iso}.

\subsection{Infinite vs finite state spaces}

Certain types of circuits, in particular single-electron devices, are such that
the number of charges in a given conductor can only take few distinct values.
They can therefore be described by truncating their infinite state space
$\{\bm q\}$ to the finite set of states relevant for the dynamics.
This will not compromise thermodynamic consistency.
Such approaches can also be used to model single-electron traps which can
be useful to model random telegraphic and $1/f$ noise, as we will discuss
in section \ref{sec:basic_MOS}.

For many other circuits, such truncation to a small state space is not possible.
Infinite state spaces are indeed a crucial ingredient of devices displaying a macroscopic limit.
This is the case of CMOS circuits for instance, where as the typical size of the
transistors is increased, the number of electrons in each conductor becomes very large.
In this case the stochastic dynamics gives rise to a deterministic non-linear dynamics
(described by Eq. \eqref{eq:deterministic}), which enables the emergence of complex phenomena
such as bistabilities, oscillations, and chaos.
Studying fluctuations can become very expensive numerically and many techniques
suitable for finite state space (e.g spectral methods) will not be applicable anymore.
Also, common approximation techniques such as the second order truncation of the Kramers-Moyal
expansion of Eq. \eqref{eq:master_eq} leading to Fokker-Planck or Langevin
equations, are known to produce incorrect results \cite{hanggi1988, horowitz2015}.
One must therefore resort to more elaborate methods such as path integrals and
large deviations techniques \cite{doi1976, lefevre2007, touchette2009}.
Such methods have been used to study stochastic chemical reaction networks
\cite{lazarescu2019} where similar problems are encountered \cite{horowitz2015}.

\section{Models for Devices}

\subsection{I-V curve characterization}
\label{sec:IV_curve}

\begin{figure}
\includegraphics[scale=.2]{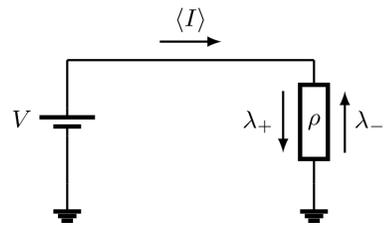}
\caption{I-V characterization of a two-terminal device.}
\label{fig:iv_curve}
\end{figure}

Two-terminal devices are usually characterized by measuring how the average electric
current through them depends on the applied voltage across their terminals, as in
Figure \ref{fig:iv_curve}. Modeling a device $\rho$ as a BPP with rates $\lambda_+$ and $\lambda_-$, and applying the LDB condition of
Eq. \eqref{eq:ldb} to this simple case, we obtain:
\begin{equation}
\log \frac{\lambda_+(V)}{\lambda_-(V)} = \beta q_e V.
\label{eq:ldb_iv}
\end{equation}
The net amount of charge going through the device in the forward direction
between time $t$ and $t+\Delta t$ is:
\begin{equation}
q(\Delta t) = q_e (N_+(\Delta t) - N_{-}(\Delta t)),
\end{equation}
where $N_{\pm}(\Delta t)$ are independent Poisson processes with
rates $\lambda_{\pm}(V)$. The average current is then:
\begin{equation}
\begin{split}
\mean{I} = \mean{q(\Delta t)/\Delta t} &= q_e (\lambda_+(V) - \lambda_{-}(V))\\
&= q_e \lambda_+(V) (1-e^{-\beta q_e V}).
\label{eq:mean_iv}
\end{split}
\end{equation}
Thus, the BPP modeling assumption and the LDB condition allows to
determine the rates $\lambda_\pm(V)$ from the measurement of the I-V curve alone,
via Eqs. \eqref{eq:ldb_iv} and \eqref{eq:mean_iv}.
In turn, from these rates we can compute any statistical moment of the electric
current $I(\Delta t)= q(\Delta t)/\Delta t$.
Therefore the full statistics of the process is completely
determined by just the mean value $\mean{I(\Delta t)}$. In particular, the
second central moment is \cite{wyatt1997, wyatt1999}:
\begin{equation}
\begin{split}
\sigma^2_I(\Delta t)
&= \mean{(I(\Delta t) - \mean{I(\Delta t)})^2}\\
&= \frac{q_e^2}{\Delta t} (\lambda_+(V) + \lambda_-(V))\\
&= \frac{q_e}{\Delta t} \mean{I(\Delta t)} \coth\left(\beta q_eV/2\right),
\end{split}
\end{equation}
which at variance with the first moment $\mean{I(\Delta t)}$ depends explicitly
on the integration time $\Delta t$. This integration time is related via the
Nyquist-Shannon sampling theorem to the frequency bandwidth
$\Delta f = 1/(2\Delta t)$ of the measurement.
Thus, in the limit of large bias ($\beta q_e V \gg 1$) we obtain the usual expression for shot noise:
\begin{equation}
\begin{split}
\sigma^2_I(\Delta t)
&= 2q_e \mean{I} \Delta f.
\end{split}
\end{equation}
Then, in this context, shot noise appears as a direct consequence of the
BPP assumption and of the LDB condition. For this reason, the fluctuations
in circuits with elements that do not display shot noise cannot be faithfully
described with this formalism. In the opposite limit where thermal effects
dominate ($\beta q_e V \ll 1$), we recover the usual expression for Johnson-Nyquist
noise \cite{sarpeshkar1993, landauer1993}.

\subsection{Specific devices}

\subsubsection{Tunnel junctions}

A tunnel junction is the simplest kind of device and the one for which
the BPP model is more natural (in some regimes of operation) \cite{wasshuber2012}.
Here we will consider a tunnel junction consisting of a sufficiently small gap
between two metallic islands at room temperature, such that electrons can tunnel through the gap.
It typically displays an Ohmic I-V curve \cite{fulton1987, wasshuber2012}:
$\mean{I} = V/R_\text{TJ}$, where the tunnel junction resistance $R_\text{TJ}$
can be computed from the specific properties of the metal conductors
and the transmission coefficient of the junction.
Using Eqs. \eqref{eq:ldb_iv} and \eqref{eq:mean_iv} we obtain the rates:
\begin{equation}
\begin{split}
\lambda_+(V) &= \frac{V}{q_e R_\text{TJ}} \frac{1}{1-e^{-\beta q_e V}} \\
\lambda_-(V) &= \frac{V}{q_e R_\text{TJ}} \frac{1}{e^{\beta q_e V}-1}.
\end{split}
\label{eq:poisson_tunnel}
\end{equation}
These expressions are well defined for any positive or negative
value of the elementary charge $q_e$, and if it changes sign then the roles
$\lambda_+$ and $\lambda_-$ are just inverted. Many other different kinds
of tunnel junctions exist, which can display strongly non-linear I-V curves
depending on the spectral densities of the materials constituting the
junction (see \cite{wasshuber2012} for a quick review).

\subsubsection{Diodes}

The characteristic curve of a p-n junction diode is often modeled via the ideal
Shockley diode equation \cite{shockley1949}:
\begin{equation}
\mean{I} = I_s \left(e^{\beta q_e V} - 1\right),
\label{eq:iv_diode}
\end{equation}
where in this case $q_e$ is the positive electron charge and
$I_s > 0$ is the reversed bias saturation current. Then the Poisson rates
are given by:
\begin{equation}
\begin{split}
\lambda_-(V) &= I_s/q_e  \\
\lambda_+(V) &= (I_s/q_e) \: e^{V/V_T},
\end{split}
\label{eq:poisson_diode}
\end{equation}
where we have defined the thermal voltage
\begin{equation}
V_T = (\beta q_e)^{-1} = k_b T/q_e.
\end{equation}

\subsubsection{MOS transistors in weak inversion}
\label{sec:basic_MOS}

\begin{figure}
\centering
\includegraphics[scale=.2]{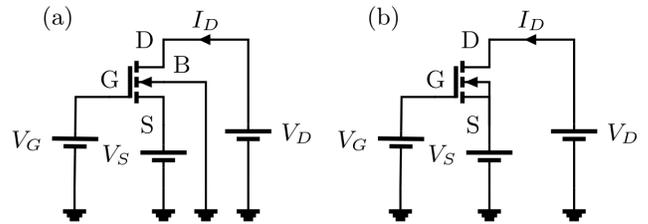}
\caption{An nMOS transistor. In (a) the bulk (B) terminal is grounded and all
other voltages are measured with respect to it. This allows to preserve the
symmetry between the source (S) and drain (D) terminals, that is broken
by connecting S and B together to obtain a three-terminal device like in (b).}
\label{fig:basic_mosfet}
\end{figure}

MOS transistors are ubiquitous devices underlying most of modern digital and analog
electronics. An \emph{enhancement-mode}
nMOS transistor like the one depicted in Figure \ref{fig:basic_mosfet}
has two typical modes of operation:
(i) a \emph{saturation} mode, and (ii) a \emph{subthreshold} or \emph{weak inversion} mode (see \cite{tsividis2011} for a rigorous discussion of all the modes of operation).
In the saturation mode the transistor essentially behaves like a switch,
allowing conduction between source (S) and drain (D) if the gate (G) voltage is
above a certain threshold $V_\text{th}$ (see Figure \ref{fig:basic_mosfet}-(a)).
If $V_G < V_\text{th}$ then source-drain conduction is suppressed. However,
whenever $V_S \neq V_D$ some small leakage current will still flow, and its magnitude
will greatly depend on how far $V_G$ is below $V_\text{th}$. This is the subthreshold
mode of operation, on which we focus in the following.
To describe this mode, we consider the Enz-Krummenacher-Vittoz model of the MOS transistor
as developed in \cite{enz2006}. According to this model, the average drain current $\meann{I_D}$
can be naturally split into forward and reverse components given by:
\begin{equation}
\begin{split}
\meann{I_D^f} &= I_0 \: e^{(V_G - V_\text{th}-\text{n} V_S)/(\text{n} V_T)}\\
\meann{I_D^r} &= I_0 \: e^{(V_G - V_\text{th}-\text{n} V_D)/(\text{n} V_T)},
\end{split}
\end{equation}
where the voltages and the current $I_D$ are defined as in Figure \ref{fig:basic_mosfet}-(a).
This model of the MOS transistor in subthreshold operation involves three
parameters characterizing the device: the threshold voltage $V_\text{th}$, the `specific'
current $I_0$ and the `slope factor' $\text{n}\geq 1$. All these parameters can be determined
from a microscopic description of the device as explained in \cite{enz2006}. The
total average drain current is then:
\begin{equation}
\begin{split}
\mean{I_D} & = \meann{I_D^f} - \meann{I_D^r} \\
&= I_0 \: e^{(V_G - V_\text{th})/(\text{n}V_T)}
(e^{-V_S/V_T} - e^{-V_D/V_T}).
\label{eq:mean_current_mosfet_bulk}
\end{split}
\end{equation}
In the previous expression the symmetry of the device is preserved, since we see
that the current $I_D$ is inverted if we interchange the roles of drain and source.
For the more common three-terminal configuration of Figure \ref{fig:basic_mosfet}-(b),
the symmetry is broken and the current is given by:
\begin{equation}
\mean{I_D} = I_0 \: e^{(V_G - V_S - V_\text{th})/(\text{n}V_T)}
(1 - e^{-(V_D-V_S)/V_T}).
\end{equation}
The voltage bias driving this current is $V_D - V_S$, which plays the role of $V$
in the I-V curve characterization.
Using this last expression we obtain the following Poisson rates:
\begin{equation}
\begin{split}
\lambda_+ &= (I_0/q_e) \: e^{(V_G - V_S - V_\text{th})/(\text{n}V_T)}\\
\lambda_- &= (I_0/q_e) \: e^{(V_G - V_S - V_\text{th})/(\text{n}V_T)}\:
e^{-(V_D- V_S)/V_T}.
\end{split}
\label{eq:poisson_mosfet}
\end{equation}
In principle this model is accurate only for $\meann{I_D^{f/r}} \ll I_0$.

\begin{figure}
\centering
\includegraphics[scale=.2]{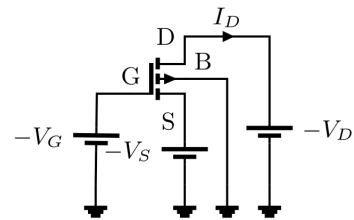}
\caption{Definition of voltage and current references for pMOS transistors.}
\label{fig:def_pfet}
\end{figure}

In a pMOS transistor conduction between drain and source is increasingly
allowed as the gate voltage becomes negative with respect to the body,
contrarily to the nMOS transistor. However, all the expressions presented for the
nMOS transistor are still valid for pMOS transistors provided that the
sign of the currents and voltages are reversed, as Figure \ref{fig:def_pfet}
indicates.

In general treatments, the noise in MOS transistors is modeled by integrating
infinitesimal Johnson-Nyquist sources along the drain-source channel
\cite{enz2006,tsividis2011}. However, for the subthreshold or weak-inversion mode
in which we are interested, the results obtained in that way are fully compatible
with those obtained from a simple BPP model as considered here \cite{sarpeshkar1993}.
This is not the case for other modes of operation.
We note that this discussion only concerns the noise
associated with the transport processes. Other kinds of noise associated with the
presence of defects and charge traps ($1/f$ and/or random telegraphic noise)
play a role in MOS transistors \cite{kirton1989, Chen2015Dec, VanBrandt2019},
and are relevant at low frequency. Random telegraphic noise can nonetheless
be modeled within the framework using individual charge traps, as briefly
mentioned in section \ref{sec:energy_diff}.
$1/f$ can be modeled as well as it can be generated by an ensemble
of random telegraphic sources \cite{paladino2014}.

\section{Charging effects}

The previously discussed I-V characterization of two-terminal devices allows to
determine the Poisson rates $\lambda_{\pm \rho}$ for a given device $\rho$
in situations where the voltage across the device is
kept fixed. In actual circuits this voltage will of course depend on the full state
$\bm{q}$ of the circuit, in accordance with the relation of Eq. \eqref{eq:capacitance_matrix}.
Thus, the Poisson rates will be functions $\lambda_{\pm \rho} (\bm{q})$ of the full
state.
However, naive constructions of the functions $\lambda_{\pm \rho} (\bm{q})$,
based on the I-V characterization and Eq. \eqref{eq:capacitance_matrix}, fail to fulfill the
LDB conditions. As a consequence, they would lead to unphysical non-thermal stationary states for closed and isothermal circuits.
To discuss and illustrate this situation, we revisit the well known
`Brillouin's paradox' \cite{brillouin1950}.
Based on the analysis of this problem we derive a procedure
to construct the rates $\lambda_{\pm \rho} (\bm{q})$ for arbitrary devices and circuits
in such a way that the LDB conditions are always respected.

\subsection{Brillouin Paradox}

\begin{figure}
\includegraphics[scale=.2]{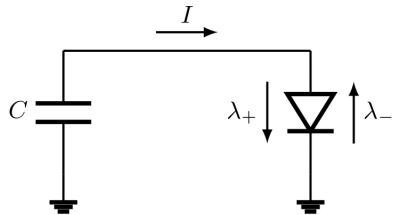}
\caption{Circuit illustrating the Brillouin's paradox.}
\label{fig:brillouin}
\end{figure}

We consider the circuit of Figure \ref{fig:brillouin}: a closed and isothermal
circuit consisting of a diode and a capacitor connected in parallel.
At any given time the voltage across the diode is equal to the capacitor
voltage $V=q/C$, where $q$ is the total charge in the upper capacitor plate.
Then, to construct the Poisson rates $\lambda_\pm(q)$ we might consider the following
procedure: to obtain the rate for a transition $q\to q\mp q_e$,
we evaluate the fixed-voltage rates $\lambda_\pm(V)$ of Eqs. \eqref{eq:poisson_diode}
at the voltage \emph{preceding} the transition (we refer to this as a `naive causality' assumption).
In this way, we have:
\begin{equation}
\begin{split}
q\to q+q_e:\qquad \lambda^\text{nc}_-(q) &= I_s/q_e  \\
q\to q-q_e:\qquad \lambda^\text{nc}_+(q) &= (I_s/q_e) \: e^{q/(C V_T)}.
\end{split}
\label{eq:poisson_diode_causal}
\end{equation}
However, these rates do not fulfill the LDB condition of Eq. \eqref{eq:ldb_closed_iso},
that for this simple case reads:
\begin{equation}
\begin{split}
\log \frac{\lambda_+(q+q_e)}{\lambda_-(q)}  &= -\beta( E(q) - E(q+q_e))\\
& = \beta q_e (V + q_e/(2C))\\
& = (q + q_e/2)/(C V_T),
\end{split}
\label{eq:ldb_brillouin}
\end{equation}
where $E(q) = q^2/(2C)$ is the energy of the circuit. As
a consequence, the stationary distribution corresponding to the transition
rates $\lambda_\pm^c(q)$ is:
\begin{equation}
P^c_\text{st}(q) \propto e^{-\frac{\beta}{2C} (q^2 + q_e q)},
\end{equation}
which deviates from the correct Gibbs equilibrium by a factor $e^{-\beta q_e q/(2C)}$.
Since this factor is an uneven function of the charge, it follows that the
stationary mean value of the charge in the capacitor is strictly below $0$.
If this were the case, the capacitor could be disconnected
from the diode and employed as a source of energy, and this process
could in principle be repeated indefinitely. This apparent violation of the
second law is essentially the Brillouin paradox, and can be considered the electronic analogue
of a Brownian ratchet.

A way to solve this problem is to notice that the LDB condition of
Eq. \eqref{eq:ldb_brillouin} would be fulfilled if the fixed-voltage rates
$\lambda_\pm(V)$ were evaluated not at the voltage before each transition,
but at the average of the voltage before and after the transition. Using
this midpoint rule we obtain the rates \cite{wyatt1997, wyatt1999}:
\begin{equation}
\begin{split}
q\to q+q_e:\:\:\: \lambda_-(q) &= I_s/q_e  \\
q\to q-q_e:\:\:\: \lambda_+(q) &= (I_s/q_e) \: e^{(q-q_e/2)/(C V_T)},
\end{split}
\label{eq:poisson_diode_midpoint}
\end{equation}
which lead to the correct Gibbs equilibrium. Later we will show that the midpoint
rule is valid in general. This means that it can be applied to the fixed-voltage
rates $\lambda_{\pm\rho}(V)$ of an arbitrary device $\rho$ to obtain
thermodynamically consistent transition rates $\lambda_{\pm\rho}(\bm{q})$ when this
device is embedded in an arbitrary circuit. Although this rule seems to be at
odds with the notion of causality, it is actually not: the probability of a
transition naturally depends on the final state as well as on the initial one.
For the naive notion of causality to
be preserved one should modify the characteristic I-V curve
of the device in question in a way that is context dependent. This in turn
challenges the idea of modularity, i.e., the notion that the behavior of
a device is not influenced by its environment and therefore can be plugged
in different circuits without modifying its description, which is a basic assumption in the usual
modeling of complex electronic circuits at the deterministic level.
However, modifications to the characteristic curve of a device due to
charging effects in its environment are well known in the study of single
electron devices, where the most explicit example is known as the Coulomb
blockade effect \cite{averin1986, fulton1987, devoret1990, ali1994, wasshuber2012, grabert2013, bagrets2003}.
In the following we illustrate the charging effects in the context of the Brillouin paradox.

\subsection{Charging effects in the I-V curve}
\label{sec:charging_effects}

\begin{figure}
\includegraphics[scale=.26]{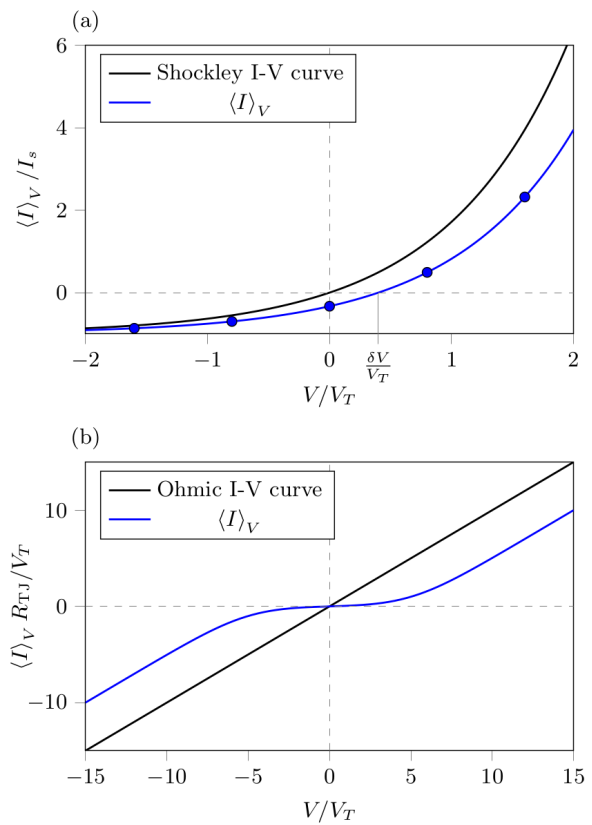}
\caption{Charging effects in the characteristic I-V curve of
 a: (a) Shockley diode ($\delta V/V_T = 0.4$), and (b) tunnel junction ($\delta V/V_T = 5$).}
\label{fig:charging_effects}
\end{figure}

Let $\mean{I}_V$ be the average current for a given value of the capacitor
voltage in the example of Figure \ref{fig:brillouin}. According to the correct rates
of Eqs. \eqref{eq:poisson_diode_midpoint}, it reads:
\begin{equation}
\mean{I}_V = I_s \left(e^{(V-q_e/(2C))/V_T} - 1 \right),
\label{eq:iv_diode_charging}
\end{equation}
which match the characteristic I-V function of
Eq. \eqref{eq:iv_diode} evaluated at a voltage shifted by $\delta V = q_e/(2C)$.
At the same time, the standard deviation of the voltage is ${\sigma_V \simeq \sqrt{k_b T/C}}$.
Then, charging effects are relevant whenever $\delta V$ is comparable to $V$ and $\sigma_V$.
Consequently, in order to observe or employ these effects, as done for example in single
electron transistors, one either needs to work with nanoscopic circuits (in order to
achieve low values of C) or at low temperatures (usually a combination of both).
The characteristic curves of Eqs. \eqref{eq:iv_diode} and \eqref{eq:iv_diode_charging}
are compared in Figure \ref{fig:charging_effects}-(a). We note that, counter to
intuition, the mean value $\mean{I}_V$ does not vanish for $V=0$. This seems
to indicate that an initially empty capacitor would charge up when connected to the
diode. However, in accordance to the second law, this anomaly is effectively
neutralized by thermal fluctuations. This can be verified by computing the mean value
of the current for the Gibbs equilibrium distribution $P_\text{eq}(q)
\propto e^{-\beta q^2/(2C)}$:
\begin{equation}
  \mean{I}_\text{eq} = \sum_q P_\text{eq}(q) \mean{I}_{V=q/C} = 0.
\end{equation}
Also, for $0<V<\delta V$ we have $\mean{I}_V < 0$, and thus it seems
that for those values of $V$ the device is actually delivering
power. However this is not the case since the actual voltage can only
take the discrete values $V = n 2\delta V$, with $n$ integer, as indicated
with dots in Figure \ref{fig:charging_effects}-(a).
Finally, we note that if we take the limit $C\to \infty$ while fixing the voltage
$V$, which correspond to a model of a perfect voltage source, charging effects disappear
(we go back to the picture of Figure \ref{fig:iv_curve}).

\subsection{General case}
\label{sec:midpoint_general}

Comparing Eqs \eqref{eq:ldb_iv} and \eqref{eq:ldb} we see that if $V$
in the fixed-voltage rates $\tilde\lambda_{\pm}(V)$
is replaced by $\mp \delta Q_{\pm\rho}(\bm{q})/q_e$,
then the resulting state dependent rates will automatically satisfy the LDB
condition (recall the definition of $\delta Q_{\pm\rho}(\bm{q})$ in Eq.
\eqref{eq:delta_Q} and note that it satisfies
${\delta Q_{-\rho} (\bm{q}+q_e\bm{\Delta}_\rho) = -\delta Q_\rho(\bm{q})}$).
Explicitly, we should consider (for $\rho>0$):
\begin{equation}
\begin{split}
	\lambda_{\pm\rho}(\bm{q}) = \tilde\lambda_{\pm}
  \left(\mp \delta Q_{\pm\rho}(\bm{q})/q_e\right).
\end{split}
\end{equation}
In turn, if device $\rho$ is connected to conductors $n$ and $m$ (with $n\to m$
as the forward direction) then we have:
\begin{equation}
\delta Q_{\pm\rho}(\bm{q}) =
\mp q_e \overline{\Delta V}^{\: \pm\rho}_{nm}(\bm{q}),
\label{eq:average_voltage}
\end{equation}
where $\overline{\Delta V}^\rho_{nm} (\bm{q})$
is the average of the voltage difference $\Delta V_{nm} = V_n - V_m$
before and after the transition ${\bm{q} \to \bm{q} + q_e \bm{\Delta}_\rho}$.
This justifies the midpoint rule mentioned above and can be easily verified
from the relation of Eq. \eqref{eq:C_matrix_split} and the definition of
$\Phi(\bm{q})$ in Eq. \eqref{eq:phi}.

Particular care should be taken for the case of the MOS transistor (or in
general with three-terminal devices that can be considered as externally
controlled two-terminal devices).
In this case the fixed-voltage transition rates $\tilde \lambda_\pm$
for the source-drain conduction do not depend only on the voltage difference
$\Delta V_{DS} = V_D - V_S$ between those terminals,
but also on the `control' voltage ${\Delta V_{GS} = V_G - V_S}$.
Generalizing the rates of Eq. \eqref{eq:poisson_mosfet} we can write:
\begin{equation}
\begin{split}
\tilde \lambda_+ (\Delta V_{GS}, \Delta V_{DS})&= f(\Delta V_{GS}) \: g_+(\Delta V_{DS})\\
\tilde \lambda_- (\Delta V_{GS}, \Delta V_{DS})&= f(\Delta V_{GS}) \: g_-(\Delta V_{DS}),
\end{split}
\label{eq:poisson_mosfet_general}
\end{equation}
where the functions $f$ and $g_\pm$ are such that the following condition is satisfied:
\begin{equation}
\log \frac{\tilde\lambda_+(\Delta V_{GS}, \Delta V_{DS})}
{\tilde\lambda_-(\Delta V_{GS},\Delta V_{DS})} = \beta q_e \Delta V_{DS}.
\end{equation}
To construct state dependent rates satisfying the LDB condition of
Eq. \eqref{eq:ldb}, we should not only replace $\Delta V_{DS}$ by its
average before and after the transition but also do the same with the
control parameter $\Delta V_{GS}$. Explicitly:
\begin{equation}
\lambda_{\pm\rho}(\bm{q}) =
\tilde\lambda_{\pm}
\left(\overline{\Delta V}_{GS}^{\pm\rho}(\bm{q}), \:\overline{\Delta V}_{DS}^{\pm\rho}({\bm{q}})\right).
\label{eq:rates_controlled_channel}
\end{equation}

\textcolor{\changescolor}{
An analysis similar to the one for the diode in section \ref{sec:charging_effects}
also holds for more general circuits, including CMOS circuits. If $C$ is the typical
capacitance at a given node of a circuit, the standard deviation of the voltage
fluctuations at that node can be estimated by its equilibrium value
$\sigma_V = \sqrt{k_bT/C} = \sqrt{V_T v_e}$,
where we have defined the elementary voltage $v_e = q_e/C$ which represents the voltage change associated to a single charge jump.
Values of $C$ as low as $50-100$ aF can be attained
in modern CMOS fabrication processes \cite{zheng2016},
that at room temperature lead to elementary voltages as high as $v_e \simeq 0.1 V_T$, and thus
$\sigma_V \simeq 0.3 V_T$. Then, we see that in modern subthreshold or near threshold
applications it is not possible to neglect charging effects nor thermal fluctuations
if the operating voltages are comparable to the thermal voltage $V_T$.
This in turn opens new possibilities that are explored in the example
of section \ref{sec:pbit}.
}

\subsection{Charging effects and non-linearity}


Circuits containing devices with non-linear I-V curves are qualitatively different
from circuits in which all devices are linear. For example, in linear RLC networks (where
the degrees of freedom are continuous and the stochastic dynamics is described by
a Fokker-Planck equation), the dynamics of the mean or expected values of voltages
and currents is decoupled from the dynamics of higher order moments, and therefore it matches the
deterministic dynamics \cite{freitas2019}. This is not anymore the case if some
non-linear element is present. However, in the kind of discrete models we are considering
here, charging effects can induce non-linear behaviors even if the characteristic
I-V curves of all the devices in the circuit are linear, as
is the case with tunnel junctions \cite{averin1986, devoret1990, ali1994}.
This is illustrated in Figure \ref{fig:charging_effects}-(b),
that was obtained in the same way as Figure
\ref{fig:charging_effects}-(a) for the diode. These induced non-linearities are a
resource that is exploited in the construction of single electron transistors
and logic gates only consisting of small conductive islands and tunnel junctions
between them \cite{wasshuber2012}. However, in the macroscopic limit in which
each conductor has many excess charges, or for high temperatures, the non-linear
effects are washed out. Consequently, in those regimes, the expected values of
voltages and currents in such circuits will obey closed and linear equations of motion
that will match the ones obtained from the deterministic description of
Eq. \eqref{eq:deterministic}. The question about the relation between the stochastic
and deterministic descriptions in general circuits is non-trivial, and will not
be addressed here in full generality. It will be analyzed for the particular example
of the CMOS inverter in Section \ref{sec:cmos_inverter}. A general treatment will be
considered elsewhere, based on Large Deviations Theory \cite{lazarescu2019, cossetto2020, freitas2021}.

\section{Stochastic thermodynamics}

So far we have established the general stochastic description of electronic circuits
and we have shown how to construct the transition rates corresponding to different
devices in a way that is thermodynamically consistent. In the following we explore
the general properties of this kind of models. We start by analyzing
the energy balance and the entropy production, i.e., we establish the first and
second laws. For this we first define the production of
heat in each device and its relation to the electric current.

\subsection{Electrical currents and heat dissipation}

Let us consider the following pair of
transitions, which are the inverse of each other:
\begin{equation}
\bm{q}\:
\xrightleftharpoons[-\rho]{\,+\rho\,} \:\bm{q}+ q_e \bm{\Delta}_\rho.
\end{equation}
The average electric current corresponding to this pair of transitions is:
\begin{equation}
\mean{I_\rho}_{\bm{q}} =
q_e (J_\rho(\bm{q},t) - J_{-\rho}(\bm{q}+q_e\bm{\Delta}_\rho,t))
 = -q_e D_{\bm{q}}^{\rho}[\bm{\mathcal{J}}].
\label{eq:average_current}
\end{equation}
And then the net average electric current corresponding to device $\rho$ is:
\begin{equation}
\mean{I_\rho} = \sum_{\bm{q}} \mean{I_\rho}_{\bm{q}}
= - q_e \sum_{\bm{q}} D_{\bm{q}}^{\rho}[\bm{\mathcal{J}}].
\end{equation}

In a similar way, the average rate at which heat is
\emph{provided by} the environment of
device $\rho$ corresponding to that pair of transitions is:
\begin{equation}
\begin{split}
\meann{\dot Q_\rho}_{\bm{q}}
&=
\delta Q_\rho(\bm{q}) \left( J_\rho(\bm{q}) -
J_{-\rho}(\bm{q}\!+\!q_e\bm{\Delta}_\rho,t)\right)\\
&=
q_e^{-1} \: \delta Q_\rho(\bm{q}) \mean{I_\rho}_{\bm{q}},
\end{split}
\end{equation}
%
%
where $\delta Q_\rho(\bm{q})$
is the change in energy of the system during transition
$\rho$ and is given by Eq. \eqref{eq:delta_Q}. Note that, as is usual in
stochastic thermodynamics but contrary to what is normally done in electronics,
heat is defined as positive when it increases the energy of the system. Recalling
Eq. \eqref{eq:average_voltage}, if device $\rho$ is connected
to conductors $n$ and $m$, we can write:
\begin{equation}
\meann{\dot Q_\rho}_{\bm{q}}
= - \overline{\Delta V}^\rho_{nm}(\bm{q}) \mean{I_\rho}_{\bm{q}}.
\end{equation}
This is the stochastic version of the usual formula for Joule heating.
Note that it is only valid at the level of transitions, and that the average
voltage difference is involved.
The net heat rate associated to device $\rho$ is:
\begin{equation}
\begin{split}
\meann{\dot Q_\rho} &= \sum_{\bm{q}}  \meann{\dot Q_\rho}_{\bm{q}}
= \sum_{\bm{q}} q_e^{-1} \: \delta Q_\rho(\bm{q}) \mean{I_\rho}_{\bm{q}}\\
&=-\sum_{\bm{q}} D_{\bm{q}}^\rho [\Phi]\: D_{\bm{q}}^\rho [\bm{\mathcal{J}}]
\: + \: \mean{I_\rho} V_r^T\bm{\Delta}_\rho^r,
\end{split}
\label{eq:diss_power}
\end{equation}
where we have employed Eq. \eqref{eq:delta_Q} to substitute $\delta Q_\rho(\bm{q})$.

\subsection{Balance of Energy}
Let us consider the rate of change in the mean value
of the potential $\Phi(\bm{q}, t)$:
\begin{equation}
d_t \mean{\Phi}
= \sum_{\bm{q}} d_t P(\bm{q}, t) \: \Phi(\bm{q}, t)
+ \partial_t \meann{\Phi}.
\end{equation}
The explicit time dependence of $\Phi$ accounts for possible external
controls of the parameters entering its definition, like the elements
of the capacitance matrix or the voltages of the regulated conductors.
Thus, the contribution $\partial_t \mean{\Phi}$ is interpreted as the rate of
work done by this external control:
\begin{equation}
\meann{\dot W_\Phi} = \partial_t \meann{\Phi}.
\label{eq:def_W_phi}
\end{equation}
Then, employing the master equation of Eq. \eqref{eq:master_eq}, we can write:
\begin{equation}
\begin{split}
d_t \meann{\Phi} - \meann{\dot W_\Phi}
&= \sum_{\rho, \bm{q}} D_{\bm{q}}^\rho [\bm{\mathcal{J}}] \: \Phi(\bm{q},t)\\
&=-\frac{1}{2} \sum_{\rho, \bm{q}} D_{\bm{q}}^\rho [\bm{\mathcal{J}}] \:
D_{\bm{q}}^\rho[\Phi]\\
&= -\sum_{\rho>0} \sum_{\bm{q}} D_{\bm{q}}^\rho [\bm{\mathcal{J}}] \:
D_{\bm{q}}^\rho[\Phi]\\
& = \sum_{\rho>0} \meann{\dot Q_\rho} - \sum_{\rho>0} \mean{I_\rho} \bm{V}_r^T \bm{\Delta}_\rho^r.
\end{split}
\end{equation}
In the second and third lines of this equation we used the
symmetry
$D_{\bm{q}}^{\rho}[\cdot] = -D_{\bm{q}+q_e \bm{\Delta}_\rho}^{-\rho}[\cdot]$.
Note that at variance with the first line,
the sums in the last line only involve positive values of $\rho$ and therefore
can be considered as sums over devices. According to Section \ref{sec:energy_diff},
the quantities
\begin{equation}
\meann{\dot W_\rho^r} = -\meann{I_\rho} \bm{V}_r^T \bm{\Delta}_\rho^r
\end{equation}
are the average rates of work performed by the voltage sources corresponding to device $\rho$.
In this way we obtain the following energy balance for a general circuit:
\begin{equation}
d_t \meann{\Phi} = \meann{\dot W_\Phi}
+ \sum_{\rho>0} \meann{\dot W_\rho^r} + \sum_{\rho>0} \meann{\dot Q_\rho}.
\label{eq:energy_balance}
\end{equation}

\subsection{Entropy production}

To the probability distribution $P(\bm{q},t)$ we assign the Shannon entropy:
\begin{equation}
\mean{S} = - k_b \sum_{\bm{q}} P(\bm{q},t) \: \log(P(\bm{q},t)),
\label{eq:ensemble_entropy}
\end{equation}
which, as the notation suggests, can be considered the average of the state dependent
entropy ${S(\bm{q},t) = -k_b \log(P(\bm{q},t))}$.
Its time derivative is
\begin{equation}
\begin{split}
d_t \mean{S} &= - k_b \sum_{\rho} \sum_{\bm{q}} D_{\bm{q}}^\rho[\bm{\mathcal{J}}]
\: \log(P(\bm{q},t)) \\
&= k_b \sum_{\rho>0} \sum_{\bm{q}} D_{\bm{q}}^\rho[\bm{\mathcal{J}}]
\:  D_{\bm{q}}^\rho[\log(P)].
\end{split}
\end{equation}
As usual, this rate of entropy change can be split into two components:
\begin{align}
 d_t \! \mean{S} &= k_b \! \sum_{\rho>0} \! \sum_{\bm{q}} D_{\bm{q}}^\rho[\bm{\mathcal{J}}]
D_{\bm{q}}^\rho[\log(P)] \\
& = k_b \! \sum_{\rho>0} \! \sum_{\bm{q}} \! \left( D_{\bm{q}}^\rho[\bm{\mathcal{J}}]
 D_{\bm{q}}^\rho[\log(\bm{\mathcal{J}})] -
D_{\bm{q}}^\rho[\bm{\mathcal{J}}]
 D_{\bm{q}}^\rho[\log(\bm{\lambda})]\right) \nonumber.
\end{align}
Using the LDB condition in Eq. (\ref{eq:ldb_compact})
the second term can be related to the average entropy production in the environment:
\begin{equation}
\begin{split}
\meann{ \dot \Sigma_e}
&= k_b \sum_{\rho>0} \sum_{\bm{q}} D_{\bm{q}}^\rho[\bm{\mathcal{J}}]
\:  D_{\bm{q}}^\rho[\log(\bm{\lambda})]
= - k_b \sum_{\rho>0} \beta_\rho \meann{\dot Q_\rho}.
\label{eq:entropy_flux}
\end{split}
\end{equation}
Thus, combining the last two equations we obtain the following expression for
the total average irreversible entropy production:
\begin{equation}
\meann{\dot \Sigma} \equiv d_t \meann{S} + \meann{\dot \Sigma_e}
=k_b \sum_{\rho>0} \sum_{\bm{q}} \! D_{\bm{q}}^\rho[\bm{\mathcal{J}}]
\:  D_{\bm{q}}^\rho[\log(\bm{\mathcal{J}})]
\geq 0,
\label{eq:entropy_prod}
\end{equation}
which is explicitly positive. This constitutes a proof of the second
law of thermodynamics in this context. We see that the entropy production
$\meann{\dot \Sigma}$ vanishes if and only if $D_{\bm{q}}^\rho[\bm{\mathcal{J}}]=0$, i.e.,
if the state is an equilibrium one (see Eq. \eqref{eq:detailed_balance}).
The fact that $\meann{\Sigma}$ corresponds to the familiar concept of entropy production
is further justified in the following.

\subsubsection{Isothermal conditions}
\label{eq:iso_entropy_decomp}

If the temperature of all the devices is the same then we can split the total
entropy production into a potential term and a work term. To see this we combine Eqs.
\eqref{eq:entropy_flux} and \eqref{eq:energy_balance} and write:
\begin{equation}
T \meann{\dot \Sigma_e} =  \meann{\dot W_\Phi} + \sum_{\rho>0} \meann{\dot W_\rho^r}
- d_t \meann{\Phi},
\end{equation}
where $T$ is the common temperature of all devices. Then, we obtain:
\begin{equation}
\begin{split}
T \meann{\dot \Sigma}
& = 
 T d_t \mean{S} - d_t \meann{\Phi} + \meann{\dot W_\Phi} +
\sum_{\rho>0} \meann{\dot W_\rho^r}\\
& = - d_t \meann{F} + \meann{\dot W_\Phi} + \sum_{\rho>0} \meann{\dot W_\rho^r},
\end{split}
\label{eq:ep_iso_split}
\end{equation}
where we have defined the average \emph{free energy} as:
\begin{equation}
\meann{F} = \meann{\Phi} - T \mean{S}.
\end{equation}

Thus, the temperature times the
total entropy production rate was expressed as a change in
a thermodynamic potential plus the total work performed on the system.
Integrating Eq. \eqref{eq:ep_iso_split}, and using the fact that $\meann{\dot \Sigma} >0$,
we recover the usual statement of the second law:
the amount of work that can be extracted from the system during an
arbitrary transformation is limited by minus the free energy difference.

\subsection{Equilibrium states}

Due to the LDB conditions, for time-independent,
closed and isothermal systems there is always
an equilibrium state satisfying Eq. \eqref{eq:detailed_balance}, and it is given
by Eq. \eqref{eq:gibbs_state}. However, equilibrium states could also exists
under more general conditions. For example, if an isothermal
and closed circuit is opened by connecting
some or all connected components to a voltage source (in such a way that
there is only one voltage source connected to each component), then
there also exists an equilibrium state, in which no currents flow. To
see this, we notice that in such a case also the second term
in Eq. \eqref{eq:delta_Q} is conservative.
In fact, we can write:
\begin{equation}
\begin{split}
	\delta Q_\rho(\bm{q})
  &= D^{\rho}_{\bm{q}}[\Phi] + q_e \bm{V}_r^T\bm{\Delta}_\rho^r\\
  &= D^{\rho}_{\bm{q}}\left[\Phi - \textstyle\sum_{n_p} V_{n_p} \:
  L_{n_p}\right],\\
\end{split}
\end{equation}
where $n_p$ indexes the regulated conductors,
$V_{n_p}$ is the voltage of conductor $n_p$, and
\begin{equation}
L_{n_p}(\bm{q}) = \bm{l'}^T_{n_p}\bm{q}
\label{eq:charge_free_conductors}
\end{equation}
is the total charge of the free conductors in the connected component
to which conductor $n_p$ belongs.
Here the vectors $\bm{l'}_{n_p}$ can be constructed as the reduction to the space of free conductors of the left eigenvectors of the incidence matrix
$\bm{\Delta}$ of the \emph{full} circuit, including the regulated conductors.
We see then that in this case the energy change during a transition
can be expressed as the change
in a state function $\Psi(\bm{q})$:
\begin{equation}
\delta Q_\rho(\bm{q}) = D_{\bm{q}}^{\rho}[\Psi],
\end{equation}
with
\begin{equation}
\Psi(\bm{q}) = \Phi(\bm{q}) - \textstyle\sum_{n_p} V_{n_p} \: L_{n_p}(\bm{q}).
\label{eq:gen_potential}
\end{equation}
Thus, it follows that for isothermal conditions
there exists an equilibrium state satisfying the global
detailed balance conditions of Eq. \eqref{eq:detailed_balance}.
It is given by:
\begin{equation}
P_\text{eq}(\bm{q}) = Z^{-1} e^{-\beta \Psi(\bm{q})} \prod_{\nu_c}
\delta [L_{\nu_c}(\bm{q}), L_{\nu_c}(\bm{q}^{(i)})],
\label{eq:gen_equil}
\end{equation}
where the index $\nu_c$ runs over the closed connected components of the circuit,
$L_{\nu_c}(\bm{q})$ is the total charge of component $\nu_c$ as defined by
Eq. \eqref{eq:conserved_quantities}, and $\bm{q}^{(i)}$ is the initial state.
As with Eq. \eqref{eq:gibbs_state}, the partition function $Z$ is such
that $P_\text{eq}(\bm{q},t)$ is normalized
and thus depends on the conserved quantities
$\{L_{\nu_c}(\bm{q}^{(i)})\}$.

\subsection{Fundamental non-equilibrium forces}
\label{sec:fund_noneq_forces}

Based on the previous discussion, we can now split the work performed by
the sources into conservative and non-conservative contributions. For this
we first split the set of regulated conductors into two categories. For each
of the open connected components in the full circuit we arbitrarily select
one of its regulated conductors. The conductors selected in this way are indexed
by $n_p = 1, \cdots, N_p$. As will be clear later, the subindex $p$ stands
for `potential'. The total number $N_p$ of `potential' conductors can be easily
seen to match the number of \emph{broken conservation laws} as defined in
section \ref{sec:cycles_and_cons_laws}. The rest of the regulated conductors
are indexed by $n_f = 1,\cdots,N_f$. In this case the subindex $f$
stands for `force', and $N_f = N-N_p$ equals the number of \emph{emergent cycles}.
In this way, to each transition $\rho$ involving a regulated conductor we can
assign:
(i) the voltage $V_{n_r(\rho)}$ of the regulated conductor
involved in that transition, and
(ii) a reference voltage $V_{n_p(\rho)}$, that is the voltage of
the regulated conductor $n_p$ that was selected as `potential'
in the corresponding open connected component.
Thus, the energy change during a transition $\rho$
involving a regulated conductor can be rewritten as:
\begin{align}
  \begin{split}
	&\delta Q_{\rho}(\bm{q})= D^{\rho}_{\bm{q}}[\Phi] + q_e \bm{V}_r^T\bm{\Delta}_{\rho}^r
  \label{eq:delta_Q_Psi}
  \end{split}\\
  \begin{split}
  &= D^{\rho}_{\bm{q}}\!\!\left[\Phi \!-\! \textstyle\sum_{n_p}\!\! V_{n_p} L_{n_p}\right]
  + q_e \: (\bm{\Delta}_\rho^r)_{n_r(\rho)} \left(V_{n_r(\rho)} \!-\! V_{n_p(\rho)}\right)\nonumber,
  \end{split}
\end{align}
where $L_{n_p}(\bm{q})$ is the total charge on the free conductors in the
open connected component of conductor $n_p$, as defined in
Eq. \eqref{eq:charge_free_conductors}. Thus, the heat rate of
device $\rho$, Eq. \eqref{eq:diss_power}, can also be expressed as:
\begin{align}
\begin{split}
&\meann{\dot Q_\rho}
= \sum_{\bm{q}} \frac{1}{q_e} \: \delta Q_\rho(\bm{q}) \mean{I_\rho}_{\bm{q}}
\label{eq:average_power}
\end{split}\\
\begin{split}
&=\!-\!\sum_{\bm{q}} D_{\bm{q}}^\rho [\Psi]\: D_{\bm{q}}^\rho [\bm{\mathcal{J}}]
\: + \: \mean{I_\rho}  (\bm{\Delta}_\rho^r)_{n_r(\rho)} \left(V_{n_r(\rho)} \!-\! V_{n_p(\rho)}\right),\nonumber
\end{split}
\end{align}
where $\Psi$ is the potential defined in Eq. \eqref{eq:gen_potential}.
To each device $\rho$ we can assign a voltage difference
${\Delta V_\rho = - (\bm{\Delta}^r_\rho)_{n_r(\rho)} \left(V_{n_r(\rho)} \!-\! V_{n_p(\rho)}\right)}$.
The minus sign in this definition was introduced in order to make $\Delta V_\rho$ positive whenever
the reference voltage is the lowest one in each connected component, and the forward direction
of device $\rho$ is the one leaving the regulated conductor. For a ``internal'' device not connected
to any regulated conductor, or for a device connected to a regulated conductor at the reference
voltage, $\Delta V_\rho = 0$. Then, we see that at most $N_f$ voltage differences $\Delta V_{\rho}$
can be different from zero. They are considered elements of a set $\{\Delta V_{n_f}\}_{n_f=1,\cdots,N_f}$
of fundamental voltage differences, or non-equilibrium forces.
Therefore, the total heat rate can be written as:
\begin{equation}
\meann{\dot Q} = -\sum_{\rho>0} \sum_{\bm{q}}
D_{\bm{q}}^\rho [\Psi]\: D_{\bm{q}}^\rho [\bm{\mathcal{J}}]
-\sum_{n_f=1}^{N_f} \mean{I_{n_f}} \Delta V_{n_f},
\label{eq:total_power_fund}
\end{equation}
where $\Delta V_{n_f}$ is one of the $N_f$ fundamental non-equilibrium forces
or voltage differences, and $\mean{I_{n_f}}$ its associated electric current.

\subsection{Minimal decomposition of the isothermal entropy production}

For isothermal settings, the entropy production can be decomposed in a similar
way as in section \ref{eq:iso_entropy_decomp}, this time in terms of the
\emph{grand potential}
\begin{equation}
\meann{\Omega} = \meann{\Psi} - T\mean{S}
\label{eq:grand_potential}
\end{equation}
and the fundamental forces $\Delta V_{n_f}$. To see this, we start by computing
the change in the potential $\Psi$:
\begin{equation}
\begin{split}
d_t \meann{\Psi}  - \partial_t \meann{\Psi} &=
\sum_{\bm{q}} d_t P(\bm{q},t) \: \Psi(\bm{q},t) \\
& = -\sum_{\rho>0} \sum_{\bm{q}}
 D_{\bm{q}}^\rho [\bm{\mathcal{J}}] \:D_{\bm{q}}^\rho [\Psi].
\end{split}
\end{equation}
As before, we define a rate of work associated to the external control of
the system:
\begin{equation}
\meann{\dot W_\Psi}  = \partial_t \meann{\Psi}.
\label{eq:work_rate}
\end{equation}
Note that this rate of work only coincides with $\meann{\dot W_\Phi}$ in
Eq. \eqref{eq:def_W_phi} if the voltages of the regulated conductors are
time independent. Now, combining the last two equations
with Eq. \eqref{eq:total_power_fund} and recalling that for isothermal
settings we have $T \meann{\dot \Sigma_e} = - \meann{\dot Q}$, we obtain:
\begin{equation}
T \meann{\dot \Sigma_e} = -d_t \meann{\Psi} + \meann{\dot W_\Psi} + \sum_{n_f}
\mean{I_{n_f}} \Delta V_{n_f},
\label{eq:energy_balance_fund}
\end{equation}
that leads to the following expression for the irreversible entropy production:
\begin{equation}
T \meann{\dot \Sigma} =
-d_t \meann{\Omega} + \meann{\dot W_\Psi} + \sum_{n_f} \meann{\dot W_{n_f}},
\label{eq:entropy_production_fund}
\end{equation}
where
\begin{equation}
\meann{\dot W_{n_f}} = \meann{I_{n_f}} \Delta V_{n_f}
\end{equation}
is naturally defined as the work rate associated to the fundamental voltage difference
$\Delta V_{n_f}$. We see that if the system is
not driven ($\meann{\dot W_\Psi}=0$) and there are no fundamental forces
$\Delta V_{n_f}$, then $d_t \meann{\Omega} = - T\dot\Sigma \leq 0$.
Also, from the fact that the capacitance matrix $\bm{C}$ is positive definite, it follows
that the thermodynamic potential $\mean{\Omega}$ is bounded from below. Thus, when
$\meann{\dot W_\Psi}=\meann{\dot W_{n_f}} = 0$,
$\meann{\Omega}$ is a Lyapunov function that reaches a minimum at equilibrium.

\subsection{Non-equilibrium Landauer principle}

Let us consider a transformation between two arbitrary, possibly non-equilibrium, states
$P^{(i)}(\bm{q})$ and $P^{(f)}(\bm{q})$. This transformation is driven by changing in time the
parameters of the circuit (for example the elements of the capacitance matrix or
the properties of some of the devices), and/or by modifying the voltages of the
regulated conductors. This can induce a parametric driving of the
potentials $\Phi$ and $\Psi$ (and thus also the free energies $F$ and $\Omega$),
as well as a change in the non-equilibrium forces $\Delta V_{n_f}$ to which
the system is subjected. In the following we will consider the time dependent
equilibrium state $P_\text{eq}(\bm{q},t)$, which is just the equilibrium state of
Eq. \eqref{eq:gen_equil} corresponding to the parameters of the system at time $t$,
and that will serve as a reference state. Also,
given an arbitrary state $P(\bm{q}, t)$
(compatible with the conserved quantities $L_{\nu_c}(\bm{q^{(i)}})$), we introduce
its \emph{relative entropy} with respect to the equilibrium state
$P_\text{eq}(\bm{q}, t)$:
\begin{equation}
\mathcal{I}(t) = D(P|P_\text{eq}) = \sum_{\bm{q}} P(\bm{q},t) \log(P(\bm{q},t)
/P_\text{eq}(\bm{q},t)),
\end{equation}
which in simple terms measures how much information should be provided in order
to identify the state $P(\bm{q},t)$ starting from $P_\text{eq}(\bm{q}, t)$. It vanishes
if and only if ${P(\bm{q}, t) = P_\text{eq}(\bm{q}, t)}$, and is always positive otherwise. By
employing the explicit form of the equilibrium state $P_\text{eq}(\bm{q}, t)$
it is easy to see that the relative entropy can be computed as a difference
between average free energies:
\begin{equation}
k_b T \: \mathcal{I}(t) = \mean{\Omega(t)} - \mean{\Omega(t)}_\text{eq},
\label{eq:free_energy_diff}
\end{equation}
where $\mean{\Omega(t)} = \sum_{\bm{q}} P(\bm{q},t)\Omega(\bm{q},t)$ is the
non-equilibrium free energy and $\mean{\Omega}_\text{eq} = \sum_{\bm{q}}
P_\text{eq}(\bm{q},t) \Omega(\bm{q},t) = -k_b T \log(Z(t))$ is the equilibrium one.
Using Eq. \eqref{eq:free_energy_diff}, we
can rewrite Eq. \eqref{eq:entropy_production_fund} as:
\begin{equation}
\meann{\dot W_\Psi}
+\sum_{n_f}\meann{\dot W_{n_f}} =
k_b T \: d_t \mathcal{I} + d_t \meann{\Omega}_\text{eq} + T \meann{\dot \Sigma}.
\end{equation}
Integrating this relation over time and using that
${\meann{\Sigma} \! = \! \int \! \meann{\dot \Sigma} dt \geq 0}$ we
obtain
\begin{equation}
\meann{W_\Psi}
+\sum_{n_f}\meann{W_{n_f}} \geq
k_b T \Delta \mathcal{I}+
\Delta \meann{\Omega}_\text{eq}.
\label{eq:landauer}
\end{equation}
Thus, the previous
expression provides
a bound for the amount of work necessary to perform (or that can be extracted
during) a transformation between arbitrary states. Importantly, this bound
explicitly takes into account the `information content'
of the initial and final states with respect to the `uninformative' equilibrium,
and it can be considered a general version of the Landauer principle \cite{esposito2011}.
The connection to the notion of logical or computational information
is achieved by splitting the state space $\{\bm{q}\}$
into different logical subspaces. This was done in \cite{sagawa2014},
where a bound equivalent to Eq. \eqref{eq:landauer} was derived.

The Landauer principle is commonly discussed in terms of physical memories that
represent logical values as quasiequilibrium
metastable states, of which the proper thermal
equilibrium is a mixture \cite{sagawa2012, sagawa2014, parrondo2015, riechers2018}.
However, it is important to notice that in some relevant kinds of electronic memories
logical values are represented by non-equilibrium steady states (NESSs)
that continuously produce entropy (for example, SRAM cells,
or the probabilistic bit discussed in section \ref{sec:pbit}).
Although the Landauer principle can anyway be applied to those cases,
the bound obtained from the right hand side of Eq. \eqref{eq:landauer}
does not take into account that continuous entropy production
(or ``housekeeping heat'' \cite{speck2005}), or any other additional dissipation
due to restrictions on the control parameters of the system \cite{kolchinsky2020constraints}.
Refinements of the Landauer principle based on lower bounds for the entropy production
in non-adiabatic transformations can be obtained \cite{schmiedl2007, sivak2012, diana2013, zulkowski2014, proesmans2020, nicholson2020},
but to the best of our knowledge the physics of computation with NESSs remains poorly explored.

\section{Stochastic trajectories and fluctuation theorems}
\label{sec:trajectories}

In the previous sections we have studied the average or expected values of
relevant quantities like the energy, entropy or work. In this section we turn
to a lower level description based on single trajectories in the state space
of the circuit, that will allow us to formulate different fluctuation theorems.
We closely follow the treatment in \cite{rao2018} for chemical reaction networks
and of \cite{rao2018detailed} for general Markov chains. Here we only present
the main results and the necessary definitions. Additional details about the
derivations can be found in the supplementary material.

We define a trajectory $\bm{\mathcal{Q}}_t$ as a particular realization of the
stochastic dynamics, from some initial time $\tau=0$ up to time $\tau=t$.
Thus, a particular trajectory is fully characterized by its initial state
$\bm{q}_{0}$, the set of transitions $\{\rho_l\}$ that took place
up to time $t$, and the times $\{t_l\}$ at which they occurred. The index $l$
takes the values $l=1,\cdots,N_t$, where $N_t$ is the number of transitions
up to time $t$. All this information can be encoded in the
\emph{trajectory probability current}:
\begin{equation}
j_\rho(\bm{q},t) = \sum_{l=1}^{N_t}
\: \delta[\rho,\rho_l] \: \delta[\bm{q},\bm{q}_{t_l}]
\: \delta(t-t_l),
\label{eq:trajectory_current}
\end{equation}
where $\bm{q}_t$ is the state immediately before instant $t$.
Different trajectories will occur with different probabilities.
If the evolution of the system is well described by
the master equation of Eq. \eqref{eq:master_eq}, then the probability density
$\mathcal{P}[\bm{\mathcal{Q}}_t]$
of observing trajectory $\bm{\mathcal{Q}}_t$ given that the initial
state is $\bm{q}_{0}$ satisfies:
\begin{equation}
\mathcal{P}[\bm{\mathcal{Q}}_t] =
\prod_{l=0}^{N_t}
e^{-\int_{t_l}^{t_{l+1}} \sum_\rho \! \lambda_\rho(\bm{q}_\tau, \tau) d\tau}
\prod_{l=1}^{N_t}
\lambda_{\rho_l}(\bm{q}_{t_l}, t_l),
\label{eq:traj_prob}
\end{equation}
where we have defined $t_{N_t + 1} = t$. The factors in the first product
account for the probabilities of not having any transition during the periods
$[t_l, t_{l+1})$, while the factors in the second product are proportional to
the probabilities of each of the jumps to take place. If we average Eq.
\eqref{eq:trajectory_current} over all trajectories then we recover
the probability currents of Eq. \eqref{eq:prob_currents}.

The average quantities defined in the previous sections can be easily extended
to individual trajectories. For example, the instantaneous electric current
and power of device $\rho$ are:
\begin{equation}
I_\rho(t) = -q_e \sum_{\bm{q}} D_{\bm{q}}^\rho[\bm{j}],
\label{eq:stochastic_current}
\end{equation}
and
\begin{equation}
\begin{split}
\dot Q_\rho(t) &= - \sum_{\bm{q}} \delta Q_\rho(\bm{q}) D_{\bm{q}}^\rho[\bm{j}]\\
&=-\sum_{\bm{q}} D_{\bm{q}}^\rho[\Psi] D_{\bm{q}}^\rho[\bm{j}]
- I_\rho(t) \Delta V_{n_f(\rho)}(t).
\label{eq:stochastic_power}
\end{split}
\end{equation}
These are just the stochastic versions of the average
quantities in Eqs. \eqref{eq:average_power} and \eqref{eq:average_current},
respectively, and are obtained by simply replacing the average current vector
$\bm{\mathcal{J}}(\bm{q}, t)$ by the stochastic one, $\bm{j}(\bm{q},t)$,
which is a vector function with components $\{j_\rho(\bm{q},t)\}_{\rho=\pm1,\cdots,\pm M}$.
$\Delta V_{n_f}(t)$ is one of the fundamental non-equilibrium forces defined
in Section \ref{sec:fund_noneq_forces}.

The net change of a state function $f(\bm{q},t)$ during a trajectory
can be expressed in terms of the currents $j_\rho(\bm{q},t)$ in the following way:
\begin{equation}
\begin{split}
\Delta f &= f(\bm{q}_t, t) - f({\bm{q}}_{0}, 0) \\
& = \int_{0}^t d\tau
\left\{
\partial_t f(\bm{q}_\tau, \tau)
+
\textstyle\sum_{\rho, \bm{q}}  j_\rho (\bm{q}, \tau) \: D_{\bm{q}}^\rho [f|_\tau]
\right\}.
\end{split}
\end{equation}
Applying this expression to the potential $\Psi(\bm{q},t)$ defined in
Eq. \eqref{eq:gen_potential} we can arrive
to the following energy balance for a given trajectory:
\begin{equation}
\Delta \Psi = W_\Psi + \sum_{n_f} W_{n_f} + \sum_{\rho>0} Q_\rho,
\label{eq:stochastic_energy_balance}
\end{equation}
where
\begin{equation}
W_\Psi = \int_{0}^t \! d\tau \: \partial_t \Psi(q_\tau,\tau)
\end{equation}
is the external driving work performed during the trajectory,
\begin{equation}
Q_\rho = \int_{0}^t \! d\tau \: \dot Q_\rho(\tau)
\label{eq:heat_traj}
\end{equation}
is the heat corresponding to device $\rho$, and
\begin{equation}
W_{n_f} = \int_{0}^t \! d\tau \: I_{n_f}(\tau) \Delta V_{n_f}(\tau)
\end{equation}
is the work performed by the fundamental non-equilibrium force $\Delta V_{n_f}$
($I_{n_f}$ is its associated electric current).

\subsection{Stochastic entropy and the integral fluctuation theorem}
As mentioned before, it is possible to define the entropy of a given state during a trajectory
in the following way \cite{seifert2005}:
\begin{equation}
S(\bm{q}, t) = -k_b \log(P(\bm{q},t)),
\label{eq:stochastic_entropy}
\end{equation}
where $P(\bm{q}, t)$ is the solution of the master equation
in Eq. \eqref{eq:master_eq} for a given initial distribution $P(\bm{q}, 0)$.
The entropy flow, i.e., the production of entropy in the environment during
a given trajectory is
\begin{equation}
\begin{split}
\Sigma_e(\bm{\mathcal{Q}}_t) &=
-k_b \sum_\rho \beta_\rho Q_\rho(t)\\
&=-k_b \int_{0}^t d\tau \textstyle\sum_{\rho, \bm{q}} j_\rho(\bm{q}, \tau)
D_{\bm{q}}^{\rho}[\log(\bm{\lambda})|_{\tau}].
\label{eq:traj_entropy_flow}
\end{split}
\end{equation}
Then, the total entropy production during the trajectory is:
\begin{equation}
\begin{split}
\Sigma(\bm{\mathcal{Q}}_t) =
&-k_b \log\left(\!\frac{P(\bm{q}_{t},t)}{P(\bm{q}_{0},0)}\!\right)\\
&-k_b \int_{0}^t d\tau \textstyle\sum_{\rho, \bm{q}} j_\rho(\bm{q}, \tau)
D_{\bm{q}}^{\rho}[\log(\bm{\lambda})|_{\tau}].
\end{split}
\label{eq:traj_full_entropy}
\end{equation}
It can be verified that the time derivatives of the averages $\mean{\Sigma_e}$
and $\mean{\Sigma}$ over all trajectories match the entropy production
rates $\meann{\dot\Sigma_e}$ and $\meann{\dot\Sigma}$ defined in Eqs. \eqref{eq:entropy_flux} and
\eqref{eq:entropy_prod}. Unlike its average $\meann{\Sigma}$, the entropy
production $\Sigma$ of a given trajectory is not always positive.
However, the fluctuations of $\Sigma$ are bound to satisfy a general
\emph{integral fluctuation theorem (IFT)}. This fundamental result is
expressed as the following equality:
\begin{equation}
\mean{e^{-\Sigma/k_b}} = 1,
\label{eq:entropy_ift}
\end{equation}
where the average is taken over all trajectories $\bm{\mathcal{Q}}_t$.
In simple terms, this equality states that positive values of the full
entropy production are more probable than negative ones.
Accordingly, from this result and Jensen's inequality the
usual statement of the second law follows: $\mean{\Sigma} \geq 0$.
Eq. \eqref{eq:entropy_ift} is valid for transient or steady state dynamics,
in autonomous or time-dependent circuits.

\subsection{Detailed fluctuation theorems}

Equation \eqref{eq:entropy_ift} is only one of several fluctuation theorems.
Under certain conditions, other quantities different from the full entropy
production satisfy more stringent constraints. For example, in isothermal
settings where $\beta_\rho = (k_b T)^{-1}$ for all $\rho$,
it is possible to obtain the following \emph{detailed fluctuation theorem} (DFT):
\begin{equation}
\frac{P(\{W_{n_f}\}, W_\Psi)}{P^\dagger(\{-W_{n_f}\}, -W_\Psi)} =
\exp((W_{\Psi} + \textstyle\sum_{n_f} W_{n_f} )/(k_b T)).
\label{eq:dft_iso_nc_c}
\end{equation}
In this expression, $P(\{W_{n_f}\}, W_\Psi)$ is the probability to observe the values
$\{W_{n_f}\}$ of work for each of the fundamental forces and of $W_\Psi$ for
the driving work during a \emph{forward protocol}. This protocol consists on the initialization
of the system state at $t=0$ according to the equilibrium state of Eq. \eqref{eq:gen_equil}, and its subsequent evolution according to the transition rates
$\lambda(\bm{q}, \tau)$ up to time $t$ (the explicit dependence of the rates on $\tau$ takes
into account a possible external manipulation of the circuit parameters, leading to an
inhomogeneous time evolution). Analogously, the quantity $P^\dagger(\{-W_{n_f}\}, -W_\Psi)$
is the probability to observe the work values $\{-W_{n_f}\}$ and $-W_\Psi$ during
the corresponding \emph{backward protocol}, in which the system is initialized
in a state drawn from the equilibrium distribution of Eq. \eqref{eq:gen_equil} (this time
corresponding to the circuit parameters at time $t$), and evolves according to the rates
$\lambda(\bm{q}, t-\tau)$.

The previous result holds for arbitrary times $t$, under the condition that the
initial state is an equilibrium one. A complementary relation can be obtained,
which is only valid asymptotically (that is, for long times), but irrespective of the initial
distribution (under the additional assumption that the circuit state is bounded to
a finite region of the state space). It reads:
\begin{equation}
\frac{P(\{\bar{\dot W}_{n_f}\}, \bar{ \dot W}_\Psi)}
{P^\dagger(\{-\bar{ \dot W}_{n_f}\}, -\bar{ \dot W}_\Psi)}
\simeq \exp(t (\bar{ \dot W}_\Psi +\sum_{n_f} \bar{\dot W}_{n_f})/(k_b T) ),
\label{eq:dft_average_rates}
\end{equation}
where we have defined the average work rates
$\bar{\dot W}_\Psi = t^{-1} W_\Psi$
and
$\bar{\dot W}_{n_f} = t^{-1} W_{n_f}$.
Note that in the case in which the voltage differences $\Delta V_{n_f}$ are
constant we have $W_{n_f} = \bar I_{n_f} \Delta V_{n_f}$, where
$\bar I_{n_f} = t^{-1} \int_0^t \! d\tau I_{n_f}(\tau)$ are the average associated
currents during the trajectory, and therefore Eq. \eqref{eq:dft_average_rates}
can be easily expressed in terms of the probabilities $P(\{\bar I_{n_f}\}, \bar{\dot W}_\Psi)$.
If there is no external driving of the circuit parameters,
then $P(\cdot) = P^{\dagger}(\cdot)$.

The fluctuation theorems in Eqs. \eqref{eq:dft_iso_nc_c} and \eqref{eq:dft_average_rates}
express fundamental symmetries of energy exchange processes. For example, the novel
thermodynamic uncertainty relations \cite{timpanaro2019}, or the usual Onsager relations
(as well as their non-linear extension \cite{gu2020}) can be recovered from them.
For completeness, a general proof of these and others fluctuation theorems is given in the
supplementary material.

\section{Applications}

\subsection{The CMOS inverter}
\label{sec:cmos_inverter}

\begin{figure}
\includegraphics[scale=.2]{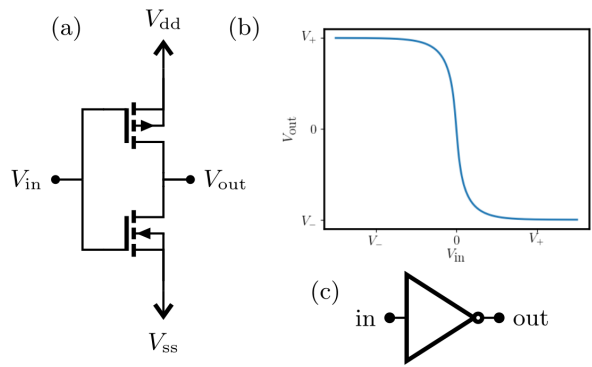}
\caption{ (a) Common implementation of a NOT gate with CMOS technology.
(b) Typical deterministic output voltage as a function of the input (for $V_\text{ss}=-V_\text{dd}$).
(c) Logical symbol for the NOT gate.}
\label{fig:inverter_diagram}
\end{figure}

The inverter or NOT gate is the most elementary logic gate. It has a single
logical input, which is negated in its only output. A diagram of a possible
implementation of this gate with MOS transistors is shown in Figure
\ref{fig:inverter_diagram}-(a). It is composed by one pMOS (top) and one
nMOS (bottom) transistors, with common drain and gate terminals.
The device is powered by applying a voltage difference $V_\text{dd} - V_\text{ss}$ between source
terminals. When the voltage in the input is $V_\text{in} < (V_\text{dd} + V_\text{ss})/2$,
conduction in the nMOS transistor is suppressed while it is enhanced in the
pMOS transistor, and therefore the output voltage $V_\text{out}$ rapidly
approaches $V_\text{dd}$. The situation is reversed for $V_\text{in} > (V_\text{dd} + V_\text{ss})/2$,
as shown in Figure \ref{fig:inverter_diagram}-(b).

\begin{figure}[ht!]
%
%
%
\includegraphics[scale=.2]{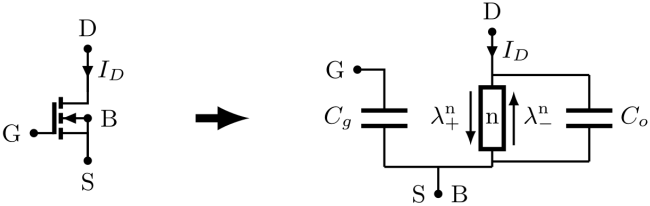}
  \caption{Model of an nMOS transistor as an externally controlled conduction
  channel between source and drain, with associated Poisson rates $\lambda^\text{n}_\pm$.
  The gate-body interface is represented as a capacitor $C_g$, and another capacitor
  $C_o$ takes into account the output capacitance. This is just a minimal model and
  does not pretend to be realistic. Other parasitic capacitances
  could also be taken into account, for example between drain and gate, but a
  proper description of them must take into account the physical
  dimensions of the device.}
  \label{fig:model_MOS}
\end{figure}
\begin{figure}[ht!]
%
%
\centering
\includegraphics[scale=.2]{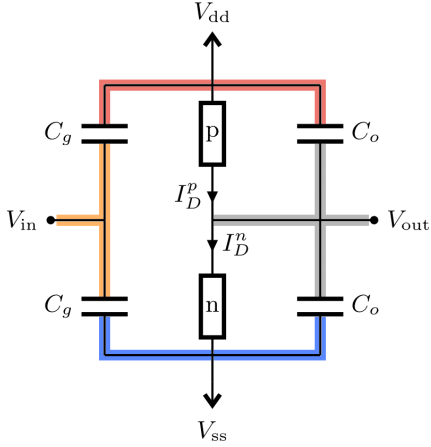}
  \caption{A possible model of the CMOS inverter. The different conductors in the circuit (regions with the same potential) are identified with different colors (see Figure \ref{fig:kinder_inverter}).}
  \label{fig:model_inverter}
\end{figure}
\begin{figure}
%
%
%
\centering
\includegraphics[scale=.2]{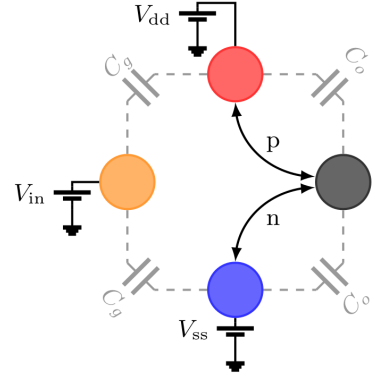}
\caption{Representation of the CMOS inverter as a set of three regulated and one free conductors,
and two conduction channels. The capacitors represent the mutual capacitances between them.}
\label{fig:kinder_inverter}
\end{figure}

\begin{figure*}[ht!]
\includegraphics[width=\textwidth]{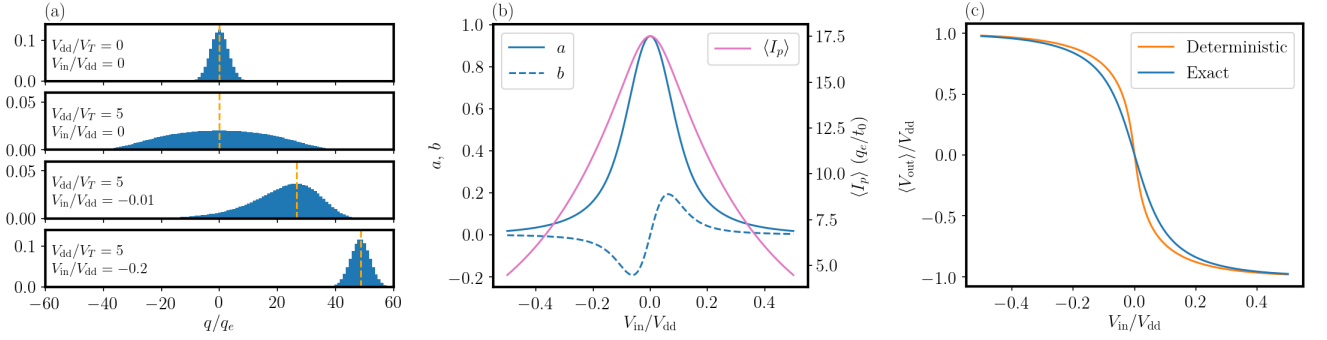}
\caption{
(a) Probability distributions for the output charge of the inverter for different power
and input voltages ($q_e/q_T = 0.1$). Here and in the other numerical results we
consider $V_\text{ss} = -V_\text{dd}$, so that $\Delta V = 2V_\text{dd}$.
The dashed lines indicate the result of
a deterministic analysis.
(b) Parameters $a$ and $b$ quantifying the deviations of
the fluctuations with respect to thermal equilibrium as a function of $V_\text{in}$, see Eq. \eqref{eq:ab_def}
($V_\text{dd}/V_T =3$ and $q_e/q_T = 0.5$). We also show the average steady state current
through the inverter (in units of $q_e/t_0$, with
$t_0 = (q_e/I_0)\exp(V_\text{th}/(\text{n}V_T))$).
(c) Comparison of the transfer function obtained from a deterministic
analysis and the exact one taking into account non-equilibrium fluctuations
($V_\text{dd}/V_T = 3$ and $q_e/q_T = 0.5$).}
\label{fig:output_inverter}
\end{figure*}

Now we explain how to build a stochastic model of the inverter within our
formalism. The first step is to model the MOS transistor as an
externally controlled conduction channel, with associated capacitances. For
example, the nMOS transistor at the bottom of the diagram of Figure
\ref{fig:inverter_diagram}-(a) can be represented as in Figure \ref{fig:model_MOS}.
There, the transistor is represented as an externally controlled conduction channel
between source and drain. The Poisson rates $\lambda^\text{n}_{\pm}$ associated to
that channel are constructed as explained in section \ref{sec:midpoint_general},
and will depend on the gate to source voltage as well as on the drain to source
voltage. The gate-body interface is modeled as a capacitor of capacitance $C_g$,
and another capacitor $C_o$ takes into account the output capacitance of the
transistor. Using this mapping, we can model the full inverter with the diagram
of Figure \ref{fig:model_inverter}. In turn, this diagram corresponds to a set of
four conductors, in which 3 of them are regulated by voltage sources, as shown
in Figure \ref{fig:kinder_inverter}.
The relation between the charges and voltages in this system is given by
\begin{equation}
\left[
\begin{matrix}
q_\text{out}\\q_\text{in}\\q_\text{dd}\\q_\text{ss}
\end{matrix}
\right]
\!
=
\!
\left[
\begin{array}{c|ccc}
2C_o & 0 & -C_o & -C_o \\
\hline
0 & 2C_g & -C_g & -C_g \\
-C_o & -C_g & C_o\! +\! C_g & 0\\
-C_o & -C_g & 0 & C_o\! +\! C_g\\
\end{array}
\right]
\!
\left[
\begin{matrix}
V_\text{out}\\V_\text{in}\\V_\text{dd}\\V_\text{ss}
\end{matrix}
\right]\!,
\end{equation}
where $q_\text{out}$ is the charge of the only free conductor (the output of the gate),
and to which we will refer simply as $q$ in the following.
By comparison with Eq. \eqref{eq:block_C_matrix} we can extract the blocks
$\bm{C}$, $\bm{C}_m$ and $\bm{C}_r$ of the capacitance matrix
(separated by lines in the previous expression). Using that and
Eq. \eqref{eq:internal_energy} we obtain the internal energy of the circuit
as a function of the only degree of freedom $q$:
\begin{equation}
E(q) \!=\! \frac{q^2}{4C_o}
\!+\! \frac{C_o}{4} \Delta V^2
\!+\! \frac{C_g}{2} \! \left[ (V_\text{in} \!-\! V_\text{dd})^2
\!+\!  (V_\text{in} \!-\! V_\text{ss})^2 \right],
\end{equation}
with $\Delta V = V_\text{dd} - V_\text{ss}$. In the same way, from Eq. \eqref{eq:phi}, we can obtain the potential
\begin{equation}
\Phi(q) = E(q) + q(V_\text{dd} + V_\text{ss})/2,
\end{equation}
and by taking its gradient we obtain the output voltage
as a function of $q$:
\begin{equation}
V_\text{out}(q) = q/(2C_o) + (V_\text{dd} + V_\text{ss})/2.
\end{equation}
Also, by selecting $V_\text{ss}$ as the reference voltage to construct the potential $\Psi$
(Eq. \eqref{eq:gen_potential}), we find
\begin{equation}
\Psi(q) = \Phi(q) - q V_\text{ss} = E(q) + q \Delta V/2,
\label{eq:psi_inverter}
\end{equation}
and that $\Delta V$ as defined above is the only non-equilibrium force (with $I_\text{p}$,
the current through the pMOS transistor, as the associated current).

According to Eq. \eqref{eq:stochastic_energy_balance}, the energy balance for
this circuit at the trajectory level is:
\begin{equation}
d_t \Psi = \dot Q_\text{n} + \dot Q_\text{p} + I_\text{p} \Delta V,
\end{equation}
where $\dot Q_\text{n}$ and $\dot Q_\text{p}$ are the heat currents associated to each of the transistors.
The irreversible entropy production is given by Eq. \eqref{eq:entropy_production_fund}
and for this case reads:
\begin{equation}
T \meann{\dot \Sigma} = -d_t \mean{\Omega} + \meann{I_\text{p}} \Delta V,
\end{equation}
where $T$ is the temperature of both transistors and
${\meann{\Omega} = \meann{\Psi} - T\meann{S}}$ (we assume time independent voltages, so the driving
contribution of Eq. \eqref{eq:work_rate} is not present).
As can be seen from the two previous equations, for steady
state conditions ($d_t \meann{\Psi} = d_t \meann{S}=0$),
the entropy production $\meann{\dot \Sigma}$ reduces to the entropy flow $\meann{\dot \Sigma_e}$
and we recover the usual expression:
\begin{equation}
T \meann{\dot \Sigma}
= -\meann{\dot Q_\text{n}} - \meann{\dot Q_\text{p}} = \meann{I_\text{p}} \Delta V .
\end{equation}

We now build the transition rates associated to both transistors according
to the procedure of Section \ref{sec:midpoint_general}. We begin with the nMOS
transistor. The voltage difference between drain and source is $\Delta V_{DS} = V_\text{out} - V_\text{ss}$,
and thus its average during the transition $q\to q\pm q_e$, with rates $\lambda^\text{n}_\mp(q)$, is
${\overline{V_{DS}} = (q\pm q_e/2)/(2C_o) + \Delta V/2}$.
For the pMOS transistor, the voltage difference between source and drain is
$\Delta V_{SD} = V_\text{dd} - V_\text{out}$ (recall that for pMOS transistors the references for
voltage and currents are reversed), and its average for the same transitions,
this time with rates $\lambda_\pm^\text{p}(q)$, is
${\overline{V_{SD}} = -(q\pm q_e/2)/(2C_o) + \Delta V/2}$. Then, via the
procedure of Section \ref{sec:midpoint_general} and the fixed-voltage rates of
Eq. \eqref{eq:poisson_mosfet}, we obtain the transition rates
\begin{equation}
\begin{split}
\lambda_+^\text{n}(q) &= (I_0/q_e) \: e^{(V_\text{in} - V_\text{ss} - V_\text{th})/(\text{n} V_T)}\\
\lambda_-^\text{n}(q) &= \lambda_+^\text{n}(q) \: e^{-((q+q_e/2)/(2C_o) + \Delta V/2 )/V_T}
\end{split}
\label{eq:rates_n_mosfet}
\end{equation}
for the nMOS transistor, and
\begin{equation}
\begin{split}
\lambda_+^\text{p}(q) &= (I_0/q_e) \: e^{(V_\text{dd} - V_\text{in} - V_\text{th})/(\text{n} V_T)}\\
\lambda_-^\text{p}(q) &= \lambda_+^\text{p}(q) \: e^{-(-(q-q_e/2)/(2C_o) + \Delta V/2 )/V_T}
\end{split}
\label{eq:rates_p_mosfet}
\end{equation}
for the pMOS. Thus, the master equation for the distribution $P(q,t)$ reads:
\begin{align}
d_t P(q, t) &= P(q-q_e, t) [\lambda_-^\text{n}(q-q_e) + \lambda_+^\text{p}(q-q_e)] \\
&+ P(q+q_e, t) [\lambda_+^\text{n}(q+q_e) + \lambda_-^\text{p}(q+q_e)] \nonumber \\
&- P(q, t) [\lambda_-^\text{n}(q) + \lambda_+^\text{n}(q) + \lambda_-^\text{p}(q) + \lambda_+^\text{p}(q)] \nonumber.
\label{eq:inverter_master_equation}
\end{align}
The master equation can be employed, for example, to find the steady state
for given voltages $V_\text{in}$, $V_\text{dd}$ and $V_\text{ss}$. As shown
in the supplementary material, this can be done analytically, and the
steady state is uniquely determined by the following recurrence relation:
\begin{equation}
P_\text{st}(q) = \frac{\alpha_\text{p} + \alpha_\text{n} \gamma e^{-(q-q_e)/q_T}}
{\alpha_\text{n} + \alpha_\text{p} \gamma e^{q/q_T}} \: P_\text{st}(q-q_e),
\label{eq:inverter_ss}
\end{equation}
where we have defined the constants:
\begin{equation}
\alpha_\text{n}  = e^{(V_\text{in} - V_\text{ss})/(\text{n} V_T)},
\qquad
\alpha_\text{p}  = e^{(V_\text{dd} - V_\text{in})/(\text{n} V_T)},
\end{equation}
and
\begin{equation}
\gamma_0 = e^{-\Delta V/(2V_T)},
\qquad\!\!\!\!
q_T = 2C_o V_T,
\qquad\!\!\!\!
\gamma = \gamma_0 e^{-q_e/(2 q_T)}.
\end{equation}
From Eq. \eqref{eq:inverter_ss} it follows that the mean value $\mean{q}_\text{st}$
can be obtained from the positive root $x$ of
\begin{equation}
\alpha_\text{p} \gamma_0 e^{a+b} \: x^2 +
 (\alpha_\text{n} - \alpha_\text{p}) \: x -
  \alpha_\text{n} \gamma_0 e^{a-b} = 0,
\label{eq:root_mean_value}
\end{equation}
as $\mean{q}_\text{st} = q_T \log(x)$. The
constants $a$ and $b$ are such that:
\begin{equation}
\mean{e^{\pm(q-\mean{q}_\text{st})/q_T}}_\text{st}
= e^{q_e/(2 q_T)} e^{a\pm b},
\label{eq:ab_def}
\end{equation}
where the mean value is taken on the stationary state given by Eq. \eqref{eq:inverter_ss}
(see supplementary material).
Thus, $a$ and $b$ quantify the fluctuations of the output charge around the mean value. They
are defined so that if the stationary state was a thermal equilibrium state (as it is for
$V_\text{dd}=V_\text{ss}$), then $a=b=0$. The constant $a$ is a measure of how the even moments of
$P_\text{st}(q)$ around the mean value deviate from those corresponding to equilibrium,
while $b$ is the same for the odd moments. We see then that Eq. \eqref{eq:root_mean_value}
determines how the non-equilibrium fluctuations, characterized by $a$ and $b$,
affect the expected output of the gate $\mean{q}_\text{st}$. If we assume that
the fluctuations are always compatible with thermal equilibrium (i.e., if $a=b=0$),
then Eq. \eqref{eq:root_mean_value} reduces to what it is obtained from a deterministic analysis of the circuit (Eq. \eqref{eq:deterministic}). Thus, this equation constitutes an exact
stochastic generalization of the deterministic results that to the best of our knowledge
was not obtained before.

In Figure \ref{fig:output_inverter}-(a) we show the probability distribution for
the output charge for different values of the power and input voltages.
When there is no voltage bias applied to the gate
($V_\text{dd} = -V_\text{ss} = 0$), the distribution is just the equilibrium one.
When a bias is applied but there is no input voltage ($V_\text{dd} = -V_\text{ss} = 5 V_T$ and $V_\text{in}=0$),
the distribution is stretched out and it ceases to be Gaussian. The application
of a small input voltage tilts this distribution to one side, and a further increase of
the input voltage generates an approximately Gaussian peak centered around the
value corresponding to the deterministic solution.
We see that the distribution of the output
charge is in general asymmetric with respect to the deterministic value. Thus, its
mean value and the deterministic one will differ. This is further evidenced
in Figure \ref{fig:output_inverter}-(b), where we plot the parameters $a$ and $b$ as a
function of the input voltage. We see that the largest deviations from equilibrium
occur around zero input voltage, when the two transistors are equally activated,
while they rapidly decrease as one of the transistors is more activated than
the other. What happens here is that for large positive (negative) $V_\text{in}$ the output
conductor is approximately at equilibrium with the source $V_\text{ss}$ ($V_\text{dd}$). Consequently, the current through the device (and therefore the entropy production), follow a similar pattern.
Finally, in Figure \ref{fig:output_inverter}-(c) we show the deviations of the
actual transfer function of the inverter from the deterministic one, caused by
non-equilibrium fluctuations.

\begin{figure}
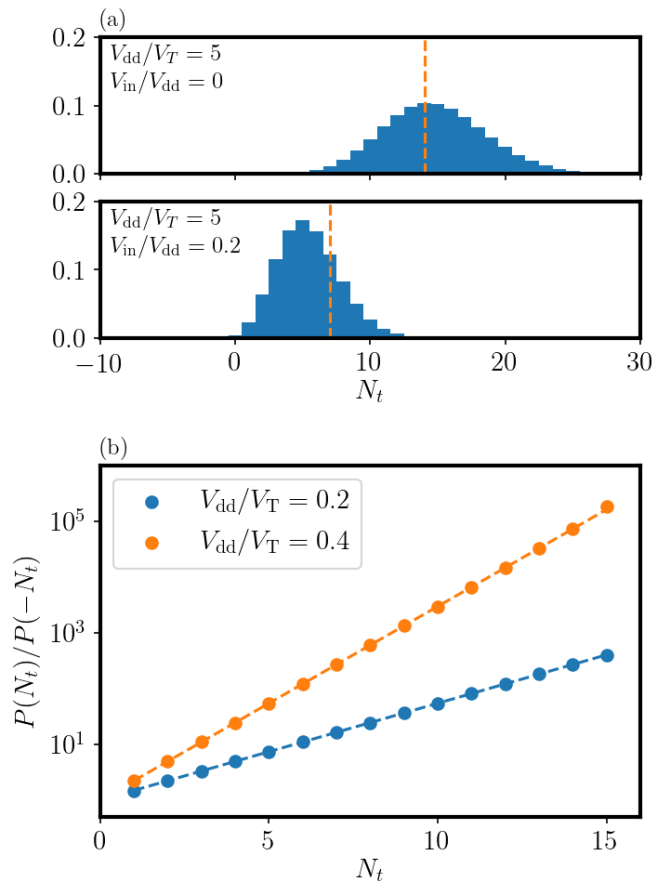

  \includegraphics[scale=.35]{\figpath{inverter_current_fluct.png}}
  \includegraphics[scale=.35]{\figpath{inverter_DFT.png}}
  \caption{(a) Probability distributions for the number of charges $N_t$ that went
  through the pMOS transistor during a time $t=10^{-1} t_0$ in the stationary
  state, for different input voltages ($q_e/q_T = 0.1$). Dashed lines
  indicate the values obtained from a deterministic analysis.
  (b) $P(N_t)/P(-N_t)$ ratio as function of $N_t$
  for two different power voltages ($q_e/q_T = 0.01$). The dots correspond
  to the numerical results obtained with the method of full counting statistics,
  and the dashed lines to the expected result according
  to the DFT of Eq. \eqref{eq:inverter_dft}.}
  \label{fig:inverter_current_fluct}
\end{figure}

We now turn to the analysis of the current fluctuations. For this we employ the
method for full counting statistics \cite{bagrets2003, esposito2007},
that we review in the supplementary material. This method
allows to evaluate the characteristic function associated to the currents fluctuations
in terms of the generator of the master equation. Then, the characteristic function
can be inverted to obtain the probability distribution. We consider
the number $N_t$ of charges that went through the pMOS transistor
during a time $t$, starting from the stationary distribution. In Figure
\ref{fig:inverter_current_fluct}-(a) we show the probability distribution of
$N_t$ for different input voltages, with $t=10^{-1}t_0$, where
$t_0 = (q_e/I_0)\exp(V_\text{th}/(\text{n}V_T))$ is the natural time scale
for this problem. Also, in Figure \ref{fig:inverter_current_fluct}-(b)
we illustrate the DFT of Eq. \eqref{eq:dft_iso_nc_c}, which in this case
reads:
\begin{equation}
\frac{P(N_t)}{P(-N_t)} = e^{N_t \Delta V/V_T},
\label{eq:inverter_dft}
\end{equation}
assuming an initial state $P_\text{eq}(q) \propto e^{-\beta\Psi(q)}$, with $\Psi(q)$
given by Eq. \eqref{eq:psi_inverter}.

From Figures \ref{fig:output_inverter}-(a) and \ref{fig:inverter_current_fluct}-(a) we
see that, except when the number of charges is too low (as in the bottom panel
of Figure \ref{fig:inverter_current_fluct}-(a)), the deterministic solution matches
the most probable result according to the stochastic treatment. This is
analogous to what has been formally shown in the case of chemical reaction
networks using large deviation theory \cite{lazarescu2019}.

These results and methods set the stage for more interesting problems,
since the NOT gate is a basic primitive in electronic design
in terms of which more complex devices can be built. For example,
connecting the output of the gate back to its input through some conduction channel
we can generate self-sustained oscillations.
Connecting two NOT gates in a loop we obtain a bistable system
with two metastable NESSs, which is the basis of many designs of electronic
memories, and also of the next example. More complex logic gates can
be modeled in the same way. We have only analyzed the stationary
distribution of the inverter for a given input voltage, although in a real
application the energetic cost of switching the inputs is a significant contribution
to the total entropy production. This cost can be analyzed by letting the input
change in time in a predefined way. Alternatively, it can also be studied
in autonomous circuits (i.e., not requiring time dependent external driving) displaying bistability
or limit cycles, as is done in the next section.

\subsection{A full-CMOS probabilistic bit}
\label{sec:pbit}

\begin{figure*}
\centering
\includegraphics[width=.95\textwidth]{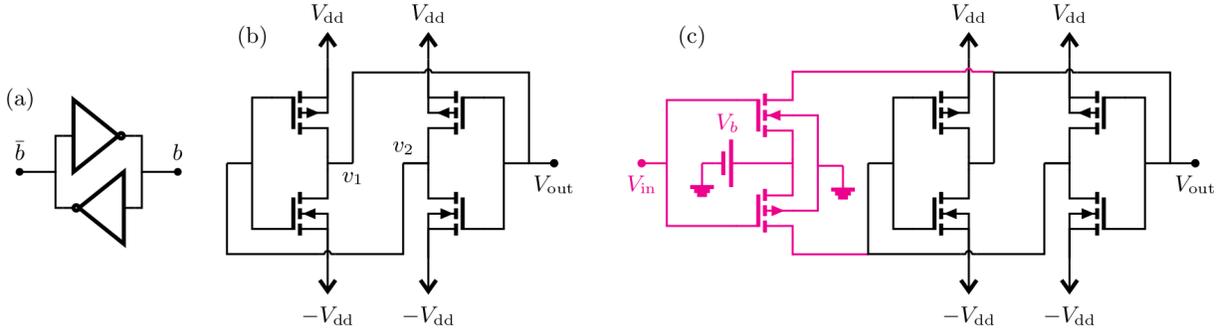}
\caption{ (a) A bistable circuit constructed with two NOT gates, representing a bit.
(b) Its CMOS implementation. (c) Complete design of a $p$-bit.
The bistable circuit constituting the bit is shown in black and is the same as in (b).
The biasing circuit is shown in magenta.}
\label{fig:design_pbit}
\end{figure*}

A probabilistic bit ($p$-bit), or binary stochastic neuron, is a device with a single output $b$
that can ideally take only two values, let us say $1$ and $-1$. It outputs the value $1$ with probability
$p$, and $-1$ with probability $1-p$. The value of $p$ is controlled by an input $I$.
For large positive values of $I$, $p\to1$, while for large negative values $p\to0$.
In a collection of $N$ of these elementary devices, the inputs $\{I_i\}_{i=1,\cdots,N}$
could be adjusted as a function of the instantaneous state $B = (b_1, b_2, \cdots, b_N)$,
and in this way correlations between different $p$-bits can be established \cite{borders2019}.
For example, given a cost function $E(B)$, it is possible to derive functional
relations $I_i(B)$ such that the state $B$ occurs with probability $P(B) \propto e^{-E(B)}$.
The function $E$ can then be chosen so that its minimum (the most probable state)
encodes the solution to some problem of interest \cite{lucas2014}. Alternatively,
the function $E(B)$ can be ``learned'' in order for $P(B)$ to
approximate the distribution of a given data set \cite{ackley1985}.

There have been recent proposals to implement $p$-bits with noisy electronic circuits.
The most relevant employs magnetic tunnel junctions (MTJs), a technology being used for some commercial memories \cite{bhatti2017},
modified on purpose so that they are sensitive to thermal noise (by lowering
the energy barrier separating the two possible states representing one bit) \cite{camsari2017, borders2019}. A previous proposal
considers a ``probabilistic switch'' \cite{cheemalavagu2005, akgul2006}: a regular CMOS inverter which is driven
by external noise at its input, so that its output fluctuates between the two
possible values. The advantage of this second proposal
is that it is based on CMOS circuits only (not on less common devices like MTJs).
However, its main drawback is that the intrinsic noise is actually neglected,
instead of being exploited as a resource. The reason is that the
description of the CMOS inverter is purely deterministic, and therefore access
to an external source of noise is assumed. Also, the noise is considered to be Gaussian,
which as we have seen in the last section is not always the case.
These limitations are of course related to the difficulty of describing
intrinsic noise in non-linear electronic circuits, as discussed in Section \ref{sec:introduction}.
As we will see next, our formalism allows to overcome those limitations,
which highlights its practical value.

\begin{figure*}
\includegraphics[width=\textwidth]{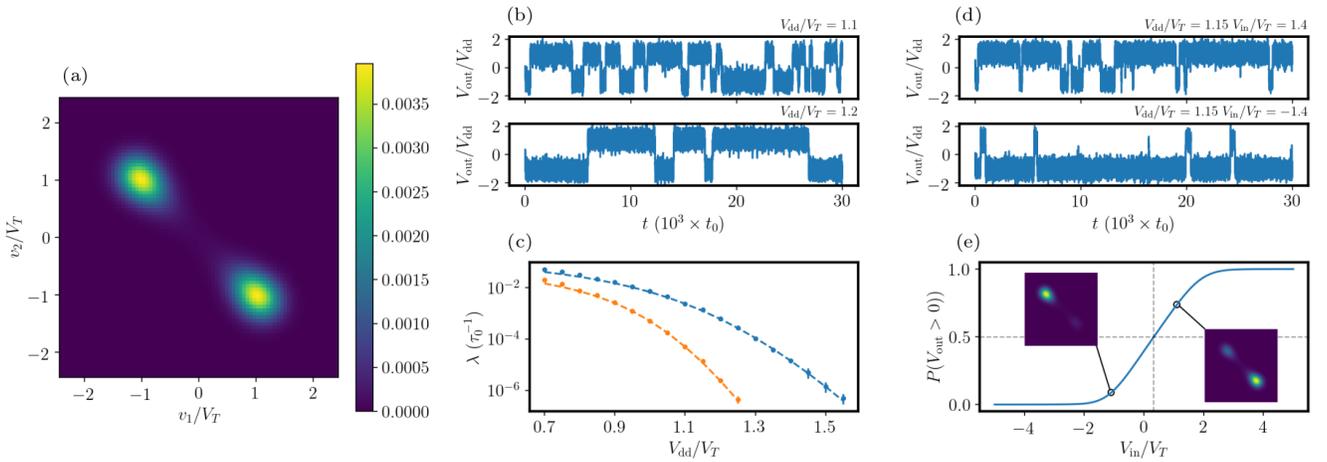}
\caption{\textcolor{\changescolor}{
(a) 2D histogram of the steady state distribution.
(b) Output voltage as a function of time for two different power voltages.
(c) Transition rate $\lambda$ as a function of the power voltage. Dashed lines correspond
to the spectral method developed in \cite{freitas2021}, and dots to the calculation of
$\lambda$ from stochastic trajectories like the ones in (b). In the last case, the error bars indicate the uncertainty in the determination of $\lambda$ ($95\%$ confidence interval), and can be reduced by generating longer trajectories. The orange line and dots
correspond to doubling the scale of the device ($C_{o,g} \to 2 C_{o,g}$).
(d) Output voltage as a function of time for two different input voltages $V_\text{in}$.
(e) Probability of the output being positive as a function of the input voltage. The insets
show the steady state distribution for $V_\text{in}/V_T = \pm1.1$. In all cases
 the parameters are ${V_T = 26 \text{ mV}}$,  $C_g = 50\text{ aF}$,  $C_o = 10^{-2} C_g$, $\text{n}=1$.}}
\label{fig:pbit_results}
\end{figure*}

We propose a full-CMOS design for a $p$-bit
that is self-sufficient: its stochastic behavior is due to the intrinsic
thermal noise, so no external source of noise is necessary.
Also, while in $p$-bits based on MTJs the transition rate or error probability
is fixed by the fabrication process (for a given temperature), our design allows
to control this parameter on the fly by just changing the power voltage.
The basic circuit is shown in Figure \ref{fig:design_pbit} and is composed of two coupled NOT gates as in regular SRAM cells.
The logical
circuit of Figure \ref{fig:design_pbit}-(a) has two stable states: $b=1$ and $\bar b=-1$,
or $b=-1$ and $\bar b=1$. The corresponding CMOS implementation
of Figure \ref{fig:design_pbit}-(b) has two degrees of freedom: the voltages $v_1$
and $v_2$ (or alternatively, the charges $q_1$ and $q_2$) at the output of each inverter.
We consider $V_\text{out} = v_1$ to be the output used to monitor the state of the bit.
If the powering voltage is above a critical value $V_\text{dd}^*$ (that for $\text{n}=1$
can be found to be $V_\text{dd}^* = V_T \ln(2)$ \cite{freitas2021}), then the
deterministic equations for the circuit have two possible steady solutions, which correspond to the two stable logical states of Figure \ref{fig:design_pbit}-(a).
At the stochastic level, they correspond to two metastable NESSs, for which $V_\text{out} \simeq V_\text{dd}$ or $V_\text{out} \simeq -V_\text{dd}$, respectively.
A stochastic model for the circuit
of Figure \ref{fig:design_pbit}-(b) can be built as before, by employing the mapping
of Figure \ref{fig:model_MOS} and constructing the rates associated to
each transistor with the procedure of Section \ref{sec:midpoint_general}.
In this case this is done automatically by a custom software package, that is
also able to deal with general circuits \cite{SSE_github}.
The steady state distribution can be obtained
by constructing the generator of the master equation in Eq. \eqref{eq:master_eq}
(truncated to some maximum number of charges),
and computing its eigenvector of zero eigenvalue, as shown in Figure \ref{fig:pbit_results}-(a).
Also, the corresponding stochastic dynamics can be simulated with the Gillespie algorithm.
In this way we can generate stochastic trajectories.
For example, in Figure \ref{fig:pbit_results}-(b)
we show two trajectories for different values
of the power voltage. To obtain those results we considered the following
parameters: $V_T = 26 \text{ mV}$ (room temperature),  $C_g = 50\text{ aF}$ and
$C_o = 10^{-2} C_g$. Crucially,
these values of capacitances are compatible to what is achieved
in modern sub-$7$nm fabrication processes \cite{zheng2016}.
Also, for simplicity we took $\text{n}=1$, and as before
the parameters $I_0$ and $V_\text{th}$ of the transistors just fix the
time scale $t_0 = (q_e/I_0)\exp(V_\text{th}/(\text{n}V_T))$.
We clearly observe random transitions, or errors, between the two metastable NESSs,
and that the transition or error rate depends on the power voltage $V_\text{dd}$.
\textcolor{\changescolor}{This is easily understood:
frequent random transitions are expected whenever
the standard deviation of the fluctuations around the output voltage,
which can be estimated as $\sigma_V = \sqrt{k_bT/(2(C_o+C_g))}$,
is comparable to the mean value $\mean{V_\text{out}} \simeq \pm V_\text{dd}$.
Thus, one can control the transition rate by changing the powering voltage
$V_\text{dd}$ or, for fixed
$V_\text{dd} > V_\text{dd}^*$, by changing the temperature and or the size of the
transistors (which modifies the capacitances $C_g$ and $C_o$).
Indeed, to a very good approximation, the waiting time $\tau$ between transitions is exponentially distributed,
$P(\tau) = \lambda e^{-\lambda \tau}$, and the transition rate
$\lambda$
can be seen to scale as
$\lambda \propto \exp\left(-2 (V_\text{dd}/\sigma_V)^2/(\text{n}+2)\right)$
to dominant order in $V_\text{dd}/\sigma_V \gg 1$ \cite{freitas2021}.
Note that for the previous parameters $V_\text{dd}^*/\sigma_V$ is of order one
at room temperature, but that it increases as the square root of the size of
the transistors. Therefore, the transition rate $\lambda$
decreases exponentially in the size of the transistors. As a consequence,
the exploitation of these naturally ocurring fluctuations as a resource
at room temperature is only a real possibility for highly scaled,
state of the art fabrication processes, as the ones considered above.
The error rate $\lambda$ can be computed from
the trajectories generated by the Gillespie algorithm,
or also by spectral methods as explained in \cite{freitas2021}.
The results are shown in Figure \ref{fig:pbit_results}-(c),
for two different scales.}

To complete the construction of the $p$-bit it is necessary to provide a mechanism
to bias its output. There are different ways to achieve this, and here we will
focus on the circuit of Figure \ref{fig:design_pbit}-(c), where the biasing circuit is
colored. It works as follows. A bias voltage $V_b>0$ is coupled to the outputs
of the two inverters which form the core of the $p$-bit through the drain-source
channel of two transistors. The transistor influencing the output of the
first inverter is an nMOS, while the one influencing the output of the
second inverter is a pMOS. Both transistors have their bodies grounded,
such that their activation depends only on the
gate-body voltage $V_\text{in}$ (the Poisson rates corresponding to this configuration
are discussed in the supplementary material). For $V_\text{in}=0$, both transistors
are equaly activated
and the output of both inverters is very weakly biased towards $V_b$ in . For $V_\text{in}>0$,
conduction through the nMOS is enhanced, while it is suppressed for the pMOS,
and therefore only the output of the first inverter is biased towards $V_b$.
In that case the symmetry between the two possible metastable NESSs ($V_\text{out}\simeq V_\text{dd}$
or $V_\text{out}\simeq -V_\text{dd}$) is broken in favour of the one with $V_\text{out}\simeq V_\text{dd}$.
The situation is reversed for $V_\text{in} <0$.
In Figure \ref{fig:pbit_results}-(d) we show two sample trajectories of the output voltage
$V_\text{out}$ for a positive and a negative value of the input voltage $V_\text{in}$.
We see that $V_\text{out}$ is indeed biased and spends more time around positive or negative
values, respectively. The parameters of the transistors are the same as before,
with the exception that the specific current $I_0'$ of the transistors in the biasing
circuit is one order of magnitude lower than the others ($I_0'=I_0/10$). Also,
we considered a bias voltage $V_b = V_T$. In Figure \ref{fig:pbit_results}-(e)
we show how the probability $p= P(V_\text{out}>0)$
of the output being positive depends on the input voltage.
We see that $p$ is indeed given by a sigmoidal function of $V_\text{in}$,
similar to the typical activation functions considered in artificial neural networks.
Note however that, due to the asymmetric I-V curves of the transistors in the biasing
circuit, the balanced case $P(V_\text{out} > 0) =1/2$ is not achieved at $V_\text{in}=0$
but for a slightly positive $V_\text{in}$.

\begin{figure*}
\centering
\includegraphics[width=\textwidth]{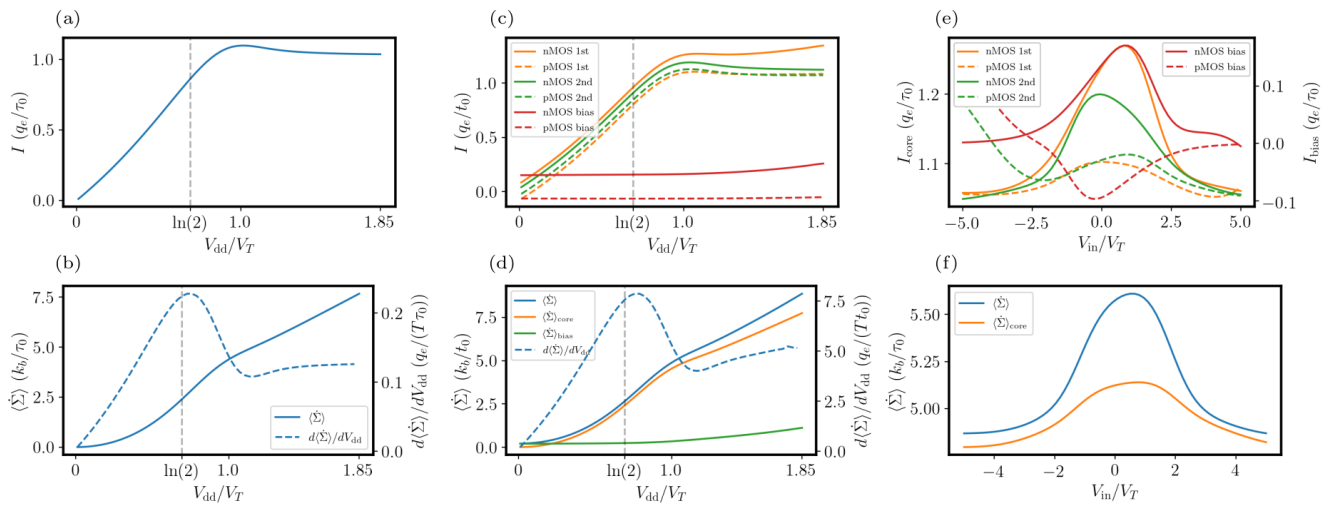}
\caption{\textcolor{\changescolor}{
For the core circuit of Figure \ref{fig:design_pbit}-(b) we show (a) the average electric current (which is the same for all the four transistors), and (b) the total entropy production rate and its derivative with respect to the powering voltage.
For the full circuit of Figure \ref{fig:design_pbit}-(c) with $V_b=V_T$ and $V_\text{in}=0$ we show (c) the average electric
current through all the transistors, and (d) the total entropy production rate and its derivative. In this case we also split the contributions to the entropy production associated to the transistors of the core and biasing circuits. In (e) and (f)
we show, respectively, the average current through all the transistors and the entropy production rate for the full circuit, this time as a function of the input voltage
for $V_\text{dd}/V_T =1.15$ and $V_b/V_T=1$. The other parameters are ${V_T = 26 \text{ mV}}$,  $C_g = 50\text{ aF}$,  $C_o = 10^{-2} C_g$, $\text{n}=1$.}}
\label{fig:pbit_currents}
\end{figure*}

\textcolor{\changescolor}{
We now analyze the energy consumption of the $p$-bit. For this we obtain the steady
state distribution as before, and compute the mean values of current and heat rates
associated to each transistor according to Eqs. \eqref{eq:average_current}
and \eqref{eq:diss_power}, respectively.
By symmetry, for the circuit consituting the core of the bit in
Figure \ref{fig:design_pbit}-(b), the steady state electric current through all the four
transistor is the same, and its dependence on $V_\text{dd}$ is shown in
Figure \ref{fig:pbit_currents}-(a). We see that it increases monotonously for low
powering voltages, and that it develops a peak right after the transition into
bistability (at $V_\text{dd} = V_T \log(2)$), after which it settles to
a constant value.  This maximum in the current close to the transition into bistability
is associated to the ocurrence of errors or transitions between different NESSs:
each switching event involves the relaxation of previously stored charge,
which adds up to the continuous flow of charge in each NESS. This is also seen
in Figure \ref{fig:pbit_currents}-(b), where we plot the entropy production rate
and its derivative with respect to the powering voltage. Since we are considering
steady state conditions, we have that
$\meann{\dot \Sigma} = \meann{\dot \Sigma_e} = (-1/T)(\meann{\dot Q_{\text{p}1}} +
\meann{\dot Q_{\text{n}1}} + \meann{\dot Q_{\text{p}2}} + \meann{\dot Q_{\text{n}2}})$,
where $\meann{\dot Q_\text{(n/p)(1/2)}}$ is the rate of heat dissipation in the
n/pMOS transistor of the first or second inverter. Interestingly, we see that
the transition to bistability is signaled by a maximum in the derivative
of $\meann{\dot \Sigma}$ with respect to $V_\text{dd}$.
}

\textcolor{\changescolor}{
In Figure \ref{fig:pbit_currents}-(c) we show all the currents for the full circuit
of Figure \ref{fig:design_pbit}-(c), including the bias circuit, for $V_b/V_T = 1$
and $V_\text{in} =0$. We see that the behaviour of the currents through the
transistor in the core circuit is qualitatively similar as the previous case,
although the currents are naturally not all equal anymore.
In Figure \ref{fig:pbit_currents}-(d) we show the total entropy production and its
derivative for the full circuit. In this case we split the total entropy production
rate as $\meann{\dot \Sigma} = \meann{\dot \Sigma}_\text{core} +
\meann{\dot \Sigma}_\text{bias}$, where
$\meann{\dot \Sigma}_\text{core} = (-1/T)(\meann{\dot Q_{\text{p}1}} +
\meann{\dot Q_{\text{n}1}} + \meann{\dot Q_{\text{p}2}} + \meann{\dot Q_{\text{n}2}})$,
and
$\meann{\dot \Sigma}_\text{bias} = (-1/T)(\meann{\dot Q_{\text{p}b}} + \meann{\dot Q_{\text{n}b}})$
are the rate entropy production rates corresponding the the bit core and biasing
circuits ($\meann{\dot Q_{(\text{n/p})b}}$ is the rate of heat dissipation in the
nMOS or pMOS biasing transistor). We see that the transition to bistability
is still signaled by a maximum in the derivative of $\meann{\dot \Sigma}$,
and that the dissipation associated to the biasing circuit is a small fraction of
the total one.
}

\textcolor{\changescolor}{
Finally, the behaviour with respect to the input voltage $V_\text{in}$ with fixed
powering voltage $V_\text{dd}/V_T = 1.15$ and bias voltage $V_b/V_T=1$
is shown in Figures \ref{fig:pbit_currents}-(e-f). In Figure
\ref{fig:pbit_currents}-(f), we see that the entropy production rate has a maximum value
at zero bias. This is again due to the ocurrence of transitions between different
NESSs, which of course decrease when the bias increases in any direction.
}

\textcolor{\changescolor}{
The average total dissipated heat per generated bit is given by
$\bar Q = T\meann{\dot \Sigma}/\lambda$. It is interesting to note that for
transitions generated by intrinsic thermal noise,
$\bar Q$ decreases as the speed or transition rate increases. For example,
by reducing the powering voltage $V_\text{dd}$ (always above $V_\text{dd}^*$),
the total rate of heat dissipation $T\meann{\dot\Sigma}$ decreases
(Figures \ref{fig:pbit_currents}-(b) and (d)),
while the transition rate $\lambda$ increases exponentially
(Figures \ref{fig:pbit_results}-(c)), and therefore $\bar Q$
decreases exponentially.
For $V_\text{dd}/V_T \simeq 1.1$ we have an average dissipated heat
per generated bit in the order of
${\bar Q = T \meann{\dot \Sigma}/\lambda \simeq 2 \times 10^{3} k_bT \simeq 10\text{ aJ} }$.
This can be compared to the MTJ $p$-bit in \cite{borders2019}, that requires
an energy of $2\text{ fJ}$ per random bit, two orders of magnitude higher
than the previous estimation.
}

\section{Discussion}

We have presented a formalism for the construction of stochastic models of non-linear
electronic circuits in a thermodynamically consistent way. Devices with arbitrary
I-V curves can be described, provided that their current fluctuations display shot noise.
A complete analysis of the stochastic thermodynamics of these models was carried out.
The relevant thermodynamic potentials were identified, and the different contributions
to the irreversible entropy production were characterized.
All these quantities were extended to individual trajectories, based on which we
presented different detailed and integral fluctuation theorems.
As a first application, we have constructed a stochastic model of a subthreshold CMOS
inverter, or NOT gate. We have shown how to analytically find the steady state of the
resulting master equation. Based on that solution, we analyzed how
the non-equilibrium thermal fluctuations induce modifications in the transfer function of the gate.
Also, we showed how to compute the full counting statistics of the current fluctuations
and in that way illustrated a detailed fluctuation theorem.
Finally, we proposed a full-CMOS design of a probabilistic bit,
or binary stochastic neuron, in which intrinsic thermal noise is exploited
as a resource to generate random bits of information in a controllable way.
The energy consumption of our design is several times lower than in previous proposals.

Of course, the formalism has some limitations, which are important to discuss here.
In first place, only devices displaying shot noise can be described. This excludes, for example,
regular resistors or MOS transistors in general modes of operation.
In addition, it is not possible to describe inertial effects (i.e., inductances). These
could be included at the price of mixing discrete and continuous variations of charge.
Nevertheless, there is little practical motivation for this, since it is difficult
to integrate inductances in nanoscale electronic circuits \cite{philip2017},
and therefore inertial effects might only be relevant at extremely high frequencies.
There are other possible extensions of the formalism, which are however less
relevant, since they capture effects that can be actually emulated with the formalism
as presented here. For example, although the stochastic dynamics we have considered is
Markovian, non-Markovian effects can be described by considering a given circuit
as a part of a larger one (something known as `Markovian embedding').
Finally, although we have not focused on single electron devices, our formalism
can be directly applied to them. In that context, there are ``cotunneling'' effects
(events in which two or more transitions happen at the same time, possibly
leaving the state of the circuit unchanged) that can become relevant in the
Coulomb blockade regime and our formalism does not take into account\cite{wasshuber2012}

Our work offers a connection between different subjects and communities.
It shows how to employ the methods of stochastic thermodynamics and single electron
devices in order to model other kinds of circuits that are traditionally given
a deterministic description, which is later supplemented by an approximate treatment of the noise.
In this way we can describe the fluctuations in those circuits
on a rigorous basis. This is relevant and timely, given the impressive
reduction in the size of CMOS circuits, and the need for new energy-efficient computing
paradigms. On the other hand, the great versatility in the fabrication and control
of electronic circuits makes them an excellent platform to study complex phenomena
in statistical physics and non-equilibrium thermodynamics. Thus, our formalism
also offers a valuable bridge between theory and experiment.
\textcolor{\changescolor}{In the future it could be interesting to explore the connection
between the low-level description we propose here and the more abstract treatments
of stochastic thermodynamics of complex circuits in Refs. \cite{boyd2018, wolpert2020},
in particular under which conditions the fundamental bounds that they obtained
can be approached within a given technology.}

After finishing this work we became aware of a recent article \cite{rezaei2020},
in which a similar stochastic description is employed to compute
the error rate of a low-power SRAM memory cell. It should be noted that the transition rates
employed in that article are actually not thermodynamically consistent, i.e., they
do not respect the local detailed balance conditions.

\section{Acknowledgments}

J. C. D. thanks Denis Flandre and Léopold Van Brandt for helpful comments and discussions.
J. C. D. and M.E. acknowledge funding from the project PDR 40003438 ``TheCirco'' on
Thermodynamics of Circuits for Computation, funded by the F.R.S.-FNRS (Belgium)
and FNR (Luxembourg).
N. F. and M. E. acknowledge funding from the European Research Council,
project NanoThermo (ERC-2015-CoGAgreement No. 681456), and from the FNR CORE program,
project NTEC (C19/MS/13664907). J. C. D. was also funded by the
FNR Program No. INTER/MOBILITY/18/12987626.

\bibliographystyle{unsrt}
\bibliography{references.bib}

\onecolumngrid


\section*{Supplementary material}

# Supplementary material for "Stochastic Thermodynamics of Non-Linear Electronic Circuits: A Realistic Framework for Computing around kT"

Nahuel Freitas,[1] Jean-Charles Delvenne,[2] and Massimiliano Esposito[1]

[1] *Complex Systems and Statistical Mechanics, Department of Physics and Materials Science, University of Luxembourg, L-1511 Luxembourg, Luxembourg*

[2] *Institute of Information and Communication Technologies, Electronics and Applied Mathematics, Université catholique de Louvain, Louvain-La-Neuve, Belgium*

PACS numbers:

## I. FLUCTUATION THEOREMS

In this section we provide additional details about the derivation of the different fluctuation theorems [1, 2]. The quantities defined in Section VI of the main text will be considered. We begin with the integral fluctuation theorem (IFT) for the total entropy production. To obtain it we need to define the concept of forward and backward protocols, and forward and backward trajectories. A *protocol* is an external manipulation of the circuit that leads to an inhomogeneous time evolution. This is reflected by the explicit time dependence on the Poisson rates $\lambda_\rho(\boldsymbol{q}, \tau)$. Given a *forward protocol* $\lambda_\rho(\boldsymbol{q}, \tau)$ up to time $t$, the corresponding *backward protocol* is given by $\lambda_\rho^\dagger(\boldsymbol{q}, \tau) = \lambda_\rho(\boldsymbol{q}, t - \tau)$. In the same way, given a *forward trajectory* $\boldsymbol{\mathcal{Q}}_t$ up to time $t$ taking values $\boldsymbol{q}_\tau$, the corresponding backward or time-reversed trajectory $\boldsymbol{\mathcal{Q}}_t^\dagger$ takes values $\boldsymbol{q}_\tau^\dagger = \boldsymbol{q}_{t-\tau}$. Note that the trajectory probability measure $\mathcal{P}[\cdot]$, defined in Eq. (94) in the main text, depends implicitly on the protocol $\lambda_\rho$. Thus, the probability density corresponding to the backward protocol $\lambda_\rho^\dagger$ is denoted as $\mathcal{P}^\dagger[\cdot]$. Using these definitions, it is possible to see that the following property holds:

$$\frac{\mathcal{P}[\boldsymbol{\mathcal{Q}}_t]}{\mathcal{P}^\dagger[\boldsymbol{\mathcal{Q}}_t^\dagger]} = \prod_{l=1}^{N_t} \frac{\lambda_{\rho_l}(\boldsymbol{q}_{t_l}, t_l)}{\lambda_{-\rho_l}(\boldsymbol{q}_{t_l} + q_e \boldsymbol{\Delta}_{\rho_l}, t_l)} = \exp\left(\Sigma_e(\boldsymbol{\mathcal{Q}}_t)/k_b\right), \quad (1)$$

since the factors corresponding to the waiting times between transitions in the forward and backward trajectories cancel out (the first factors in Eq. (94)). The last equality follows from the definition of $\Sigma_e$ in Eq. (103) in the main text. In the same way, we can write:

$$\frac{\mathcal{P}[\boldsymbol{\mathcal{Q}}_t] P(\boldsymbol{q}_0, 0)}{\mathcal{P}^\dagger[\boldsymbol{\mathcal{Q}}_t^\dagger] P(\boldsymbol{q}_t, t)} = \exp\left(\Sigma(\boldsymbol{\mathcal{Q}}_t)/k_b\right). \quad (2)$$

Rearranging the factors in the previous expression and integrating over all initial states $\boldsymbol{q}_0$ and trajectories $\boldsymbol{\mathcal{Q}}_t$, we obtain the mentioned IFT:

$$\left\langle e^{-\Sigma/k_b} \right\rangle = 1. \quad (3)$$

### A. Detailed fluctuation theorems

To derive the detailed fluctuation theorems it is convenient to introduce a particular splitting of the entropy production with respect to a, in principle arbitrary, reference state. Thus, let us consider an arbitrary probability distribution $P_{\text{ref}}(\boldsymbol{q}, t)$ over the state space, with an associated entropy and average currents:

$$S_{\text{ref}}(\boldsymbol{q}, t) = -k_b \log(P_{\text{ref}}(\boldsymbol{q}, t)), \quad (4)$$

$$J_\rho^{\text{ref}}(\boldsymbol{q}, t) = \lambda_\rho(\boldsymbol{q}, t) P_{\text{ref}}(\boldsymbol{q}, t). \quad (5)$$

Adding and subtracting the change $\Delta S_{\text{ref}}$ in the reference entropy during a trajectory to the full entropy production of Eq. (104), it is possible to arrive at the following decomposition:

$$\Sigma(\boldsymbol{\mathcal{Q}}_t) = \Sigma_{\text{nc}}(\boldsymbol{\mathcal{Q}}_t) + \Sigma_{\text{c}}(\boldsymbol{\mathcal{Q}}_t), \quad (6)$$

where

$$\Sigma_{\text{nc}}(\boldsymbol{\mathcal{Q}}_t) = -k_b \int_0^t d\tau \sum_{\rho,\boldsymbol{q}} j_\rho(\boldsymbol{q},\tau)\, D_{\boldsymbol{q}}^\rho[\log(\boldsymbol{\mathcal{J}}^{\text{ref}})|_\tau], \tag{7}$$

and

$$\begin{aligned}\Sigma_{\text{c}}(\boldsymbol{\mathcal{Q}}_t) &= \int_0^t d\tau\, \partial_t S_{\text{ref}}(\boldsymbol{q}_\tau,\tau) - \Delta D \\ &= \Sigma_{\text{d}}(\boldsymbol{\mathcal{Q}}_t) - \Delta D.\end{aligned} \tag{8}$$

In the last expression, $\Delta D$ is the change in the stochastic relative entropy between the actual distribution and the reference one, $D(t) = k_b \log(P(\boldsymbol{q}_t,t)/P_{\text{ref}}(\boldsymbol{q}_t,t))$. The subindices nc, c, and d, stand for 'non-conservative', 'conservative' and 'driving', respectively. The physical meaning of these different contributions to the entropy production will be clarified in the following.

Before proceeding it is necessary to update the previous definitions of forward and backward protocols. In the following by forward protocol we consider a process that is initiated at time $t = 0$ by selecting an initial state according to the distribution $P_{\text{ref}}(\boldsymbol{q},0)$, and subsequently evolves according to the rates $\lambda_\rho(\boldsymbol{q},\tau)$ up to time $t$. In the backward protocol, the initial state is drawn from the distribution $P_{\text{ref}}(\boldsymbol{q},t)$, and evolves with rates $\lambda_\rho^\dagger(\boldsymbol{q},\tau)$. Crucially, the time inversion is also applied to the reference distribution, i.e., during the backward trajectory the reference distribution is $P_{\text{ref}}^\dagger(\boldsymbol{q},\tau) = P_{\text{ref}}(\boldsymbol{q},t-\tau)$. Under these conditions, it can be seen that the quantities $\Sigma_{\text{nc}}$ and $\Sigma_{\text{d}}$ are odd under time inversion:

$$\Sigma_{\text{nc}}(\boldsymbol{\mathcal{Q}}_t^\dagger) = -\Sigma_{\text{nc}}(\boldsymbol{\mathcal{Q}}_t) \qquad \text{and} \qquad \Sigma_{\text{d}}(\boldsymbol{\mathcal{Q}}_t^\dagger) = -\Sigma_{\text{d}}(\boldsymbol{\mathcal{Q}}_t). \tag{9}$$

Owing to this 'involution' property, as explained in [2], the following *detailed fluctuation theorem* (DFT) holds:

$$\frac{P(\Sigma_{\text{nc}}, \Sigma_{\text{d}})}{P^\dagger(-\Sigma_{\text{nc}}, -\Sigma_{\text{d}})} = \exp((\Sigma_{\text{nc}} + \Sigma_{\text{d}})/k_b). \tag{10}$$

Here, $P(\Sigma_{\text{nc}}, \Sigma_{\text{d}})$ is the probability of obtaining the values $\Sigma_{\text{nc}}$ and $\Sigma_{\text{d}}$ of the observables $\Sigma_{\text{nc}}(\boldsymbol{\mathcal{Q}}_t)$ and $\Sigma_{\text{d}}(\boldsymbol{\mathcal{Q}}_t)$ during the forward protocol, and $P^\dagger(\Sigma_{\text{nc}}, \Sigma_{\text{d}})$ is the same for the backward protocol. Integrating the previous expression over all possible values of $\Sigma_{\text{nc}}$ and $\Sigma_{\text{d}}$ we obtain a new IFT:

$$\left\langle e^{-(\Sigma_{\text{nc}}+\Sigma_{\text{d}})/k_b} \right\rangle = 1. \tag{11}$$

Notice that this is not equivalent to the IFT of Eq. (3). Also, we note that a DFT like Eq. (10) for the full entropy production $\Sigma$ does not hold in general, due to the fact that this observable does not satisfy the involution property.

We now introduce a particular choice for the reference distribution, that naturally leads to the 'conservative' and 'non-conservative' labels introduced before. This choice is given by the equilibrium distribution of Eq. (79) in the main text, at a reference inverse temperature $\beta_{\text{ref}} = 1/(k_b T_{\text{ref}})$:

$$P_{\text{ref}}(\boldsymbol{q},t) = \frac{e^{-\beta_{\text{ref}}\Psi(\boldsymbol{q},t)}}{Z_{\text{ref}}(t)} \prod_{\nu_c} \delta[L_{\nu_c}(\boldsymbol{q}), L_{\nu_c}(\boldsymbol{q}^{(i)})]. \tag{12}$$

In this case the non-conservative contribution is

$$\Sigma_{\text{nc}} = \Sigma_e + (\sum_\rho Q_\rho + \sum_{n_f} W_{n_f})/T_{\text{ref}}, \tag{13}$$

and the driving contribution is

$$\Sigma_{\text{d}} = (W_\Psi - \Delta\langle\Omega\rangle_{\text{ref}})/T_{\text{ref}}, \tag{14}$$

where $\Omega$ is the potential defined in Eq. (83) in the main text and

$$\Delta\langle\Omega\rangle_{\text{ref}} = -k_b T_{\text{ref}} \log\left(\frac{Z_{\text{ref}}(t)}{Z_{\text{ref}}(0)}\right). \tag{15}$$

Note that this contribution to $\Sigma_d$ is not fluctuating, i.e., it does not depend on the actual trajectory. We see that if the potential $\Psi(\boldsymbol{q},t)$ is time independent, then the driving contribution vanishes. In the general case we have:

$$\Sigma_{\text{nc}} + \Sigma_{\text{d}} = \Sigma_e + (\Delta\Psi - \Delta\langle\Omega\rangle_{\text{ref}})/T_{\text{ref}}. \tag{16}$$

In isothermal conditions where $\beta_\rho = \beta_{\text{ref}} = (k_b T)^{-1}$ for all $\rho$, since $\Sigma_e = -\sum_\rho Q_\rho/T$, the non-conservative contribution is simplified to:

$$\Sigma_{\text{nc}} = \sum_{n_f} W_{n_f}/T. \tag{17}$$

Also, since the individual work contributions satisfy the involution property $W_{n_f}(\mathcal{Q}_t) = -W_{n_f}(\mathcal{Q}_t^\dagger)$, the DFT of Eq. (10) can be generalized to:

$$\frac{P(\{W_{n_f}\}, W_\Psi)}{P^\dagger(\{-W_{n_f}\}, -W_\Psi)} = \exp((W_\Psi + \sum_{n_f} W_{n_f})/(k_b T)), \tag{18}$$

where $P(\{W_{n_f}\}, W_\Psi)$ is the probability to observe the values $\{W_{n_f}\}$ of work for each of the fundamental forces and of $W_\Psi$ for the driving work during the forward protocol, while $P^\dagger(\{W_{n_f}\}, W_\Psi)$ is the same for the backward protocol.

### B. Adiabatic-Nonadiabatic decomposition

In the context of Eq. (10), another relevant option for the reference state is given by the instantaneous stationary distribution. That is, we consider the distribution $P_{\text{ref}}(\boldsymbol{q}, t)$ that is left invariant given the transition rates $\lambda_\rho(\boldsymbol{q}, t)$ for a fixed time $t$:

$$0 = \sum_\rho D_{\boldsymbol{q}}^\rho [\boldsymbol{\mathcal{J}}^{\text{ref}}|t] \qquad \forall \boldsymbol{q}. \tag{19}$$

Thus, $P_{\text{ref}}(\boldsymbol{q}, t)$ is the stationary distribution to which the system would eventually relax if the transition rates were fixed at their values at time $t$. For slowly varying protocols $d_t \langle \Sigma_{\text{nc}} \rangle$ approximates the entropy production in the instantaneous steady state, while $\langle \Sigma_c \rangle$ is associated to the driving and relaxation. Accordingly, this contributions are now refered to as 'adiabatic' and 'non-adiabatic', respectively. As explained in [1], in this case both expected values are positive: $\langle \Sigma_{\text{nc}} \rangle \geq 0$ and $\langle \Sigma_c \rangle \geq 0$.

If the instantaneous stationary distribution is an equilibrium state, and therefore satisfies the global detailed balance conditions $D_{\boldsymbol{q}}^\rho [\boldsymbol{\mathcal{J}}^{\text{ref}}] = 0$ for all $\boldsymbol{q}$ and $\rho$, then the adiabatic contribution $\Sigma_{\text{nc}}$ is strictly zero (at the trajectory level, not just on average). Thus, the DFT of Eq. (10) reduces to:

$$\frac{P(\Sigma_{\text{d}})}{P^\dagger(-\Sigma_{\text{d}})} = \exp(\Sigma_{\text{d}}/k_b). \tag{20}$$

On the other hand, if the circuit parameters do not depend on time, $\Sigma_d = 0$ and we have

$$\frac{P(\Sigma_{\text{nc}})}{P(-\Sigma_{\text{nc}})} = \exp(\Sigma_{\text{nc}}/k_b). \tag{21}$$

Note that only the probabilities for the forward protocol appear in the previous equation, since in this case the forward and backward protocols are equivalent (the starting distributions are the same, i.e., the stationary one, and there is no driving).

### C. Asymptotic average rates

In typical out of equilibrium processes the entropy production for long times is dominated by the entropy flow $\Sigma_e$, since, contrary to the change $\Delta S$ in the internal entropy, it is an unbounded and time extensive quantity (we assume that the state $\boldsymbol{q}$ dwells in a finite region of the state space). This property can be exploited to obtain a DFT for the stochastic asymptotic average work rates, or average powers. To see this we first note the following general identity:

$$\begin{aligned} \Sigma_{\text{nc}} + \Sigma_{\text{d}} &= \Sigma_e + \Delta S + \Delta D \\ &= \Sigma_e + \Delta S_{\text{ref}}, \end{aligned} \tag{22}$$

where $\Delta S_{\text{ref}} = -k_b \log(P_{\text{ref}}(\boldsymbol{q}_t, t)/P_{\text{ref}}(\boldsymbol{q}_0, 0))$. If we are considering a process in which the entropy production $\Sigma$ is time extensive, then the entropy flow $\Sigma_e$ should also be time extensive, since $\Delta S_{\text{ref}}$ is bounded and cannot grow indefinitely. Thus, recalling that $\Sigma_e = -\sum_\rho Q_\rho/T_\rho$, for long times we can write:

$$\Sigma_{\text{nc}} + \Sigma_{\text{d}} = -t \sum_\rho \bar{\dot{Q}}_\rho(t)/T_\rho + \mathcal{O}(t^0), \tag{23}$$

where

$$\bar{\dot{Q}}_\rho(t) = \frac{1}{t}\int_0^t d\tau\, \dot{Q}_\rho(\tau) \tag{24}$$

is the average heat rate of device $\rho$ during a trajectory. Thus, neglecting the non time-extensive terms in $\Sigma_{nc} + \Sigma_d$, it is possible to arrive to the following result:

$$\frac{P(\{\bar{\dot{Q}}_\rho\})}{P^\dagger(\{-\bar{\dot{Q}}_\rho\})} \simeq e^{-t\sum_\rho \beta_\rho \bar{\dot{Q}}_\rho} \tag{25}$$

for long times. This result can be further simplified for isothermal conditions, since in that case $T\Sigma_e = -Q = W_\Psi - \Delta\Psi + \sum_{n_f} W_{n_f}$. Again, assuming that the circuit state $\boldsymbol{q}$ is restricted to a finite region of the state space during the trajectory, only the quantities $W_\Psi$ and $W_{n_f}$ can be time extensive, since $\Delta\Psi$ is bounded. Thus, again neglecting non time extensive terms, for long times we can write:

$$\frac{P(\{\bar{\dot{W}}_{n_f}\}, \bar{\dot{W}}_\Psi)}{P^\dagger(\{-\bar{\dot{W}}_{n_f}\}, -\bar{\dot{W}}_\Psi)} \simeq e^{t\beta(\bar{\dot{W}}_\Psi + \sum_{n_f} \bar{\dot{W}}_{n_f})}, \tag{26}$$

where we have defined the average work rates $\bar{\dot{W}}_\Psi = t^{-1}W_\Psi$ and $\bar{\dot{W}}_{n_f} = t^{-1}W_{n_f}$.

## II. STEADY STATE OF THE CMOS INVERTER

In this section we find the steady state distribution for the NOT gate using the ladder operator method introduced in [3], where it was applied to a diode in thermal equilibrium. Thus, we define the ladder operator $X = e^{q_e d/dq}$ acting on functions over the state space:

$$Xf(q) = f(q + q_e), \qquad \text{and therefore} \qquad X^{-1}f(q) = f(q - q_e). \tag{27}$$

Then we can write the master equation for the NOT gate as

$$(q_e/I_0)\, e^{V_{\text{th}}/(nV_T)}\, d_t P(q,t) = \left\{ X^{-1}\left[\alpha_n \gamma e^{-q/(2C_oV_T)} + \alpha_p\right] + X\left[\alpha_n + \alpha_p \gamma e^{q/(2C_oV_T)}\right] \right. \\ \left. - \left[\alpha_n\left(\gamma e^{-q/(2C_oV_T)} + 1\right) + \alpha_p\left(\gamma e^{q/(2C_oV_T)} + 1\right)\right] \right\} P(q,t), \tag{28}$$

where we have defined the constants

$$\alpha_n = e^{(V_{\text{in}}-V_{\text{ss}})/(nV_T)} \qquad \alpha_p = e^{(V_{\text{dd}}-V_{\text{in}})/(nV_T)} \qquad \gamma = e^{-(\Delta V/2 + q_e/(4C_o))/V_T} \qquad q_T = 2C_oV_T. \tag{29}$$

Reorganizing the terms in Eq. (28) and using the identity $(X^{-1} - 1) = -(X-1)X^{-1}$ we obtain:

$$(q_e/I_0)\, e^{V_{\text{th}}/(nV_T)}\, d_t P(q,t) = (X-1)\left\{\alpha_n\left[1 - X^{-1}\gamma e^{-q/q_T}\right] + \alpha_p\left[\gamma e^{q/q_T} - X^{-1}\right]\right\} P(q,t). \tag{30}$$

From this, we can conclude that for the stationary state $P_{\text{st}}(q)$ the quantity

$$\left\{\alpha_n\left[1 - X^{-1}\gamma e^{-q/q_T}\right] + \alpha_p\left[\gamma e^{q/q_T} - X^{-1}\right]\right\} P_{\text{st}}(q) \tag{31}$$

should be constant. Also, this constant should be 0, since $P_{\text{st}}(q) \to 0$ for $q \to \pm\infty$. In this way we arrive at the following recurrence relation:

$$\left(\alpha_n + \alpha_p \gamma e^{q/q_T}\right) P_{\text{st}}(q) = \left(\alpha_p + \alpha_n \gamma e^{-(q-q_e)/q_T}\right) P_{\text{st}}(q - q_e), \tag{32}$$

which can be iterated in order to find $P_{\text{st}}(q)$. Its physical meaning is transparent: the left hand side is proportional to the probability of the transition $q \to q - q_e$, while the right hand side is proportional to the probability of the transition $q - q_e \to q$. The condition that these two probabilities must balance each other uniquely determines the steady state. At equilibrium it reduces to the detailed balance condition. Furthermore, by summing the previous relation over all values of $q$ we obtain:

$$\alpha_n - \alpha_n \gamma \left\langle e^{-q/q_T}\right\rangle_{\text{st}} = \alpha_p - \alpha_p \gamma \left\langle e^{q/q_T}\right\rangle_{\text{st}}, \tag{33}$$

where $\langle\cdot\rangle_{\text{st}}$ indicates averaging over the stationary distribution. This condition is just equivalent to demanding that in the stationary state the average current through both transistors must be the same. In the next section we use this relation to obtain an approximate analytical expression for the transfer function of the gate.

## III. STOCHASTIC CORRECTIONS TO THE TRANSFER FUNCTION OF THE CMOS INVERTER

The relation of Eq. (33) can be rewritten as:

$$\alpha_{\rm n} - \alpha_{\rm n}\gamma e^{-\langle q\rangle_{\rm st}/q_T} \left\langle e^{-(q-\langle q\rangle_{\rm st})/q_T} \right\rangle_{\rm st} = \alpha_{\rm p} - \alpha_{\rm p}\gamma e^{\langle q\rangle_{\rm st}/q_T} \left\langle e^{(q-\langle q\rangle_{\rm st})/q_T} \right\rangle_{\rm st}. \tag{34}$$

Now, one can notice that the quantities $\langle \exp(\pm(q-\langle q\rangle_{\rm st})/q_T)\rangle_{\rm st}$ will equal 1 in a deterministic limit. Their deviation from this value is a measure of the thermal fluctuations. If known, the previous equation could be solved to obtain $\langle q\rangle_{\rm st}$. One possible approximation is to compute them using the equilibrium distribution. Doing so, one obtains:

$$\left\langle e^{\pm(q-\langle q\rangle_{\rm eq})/q_T} \right\rangle_{\rm eq} = e^{q_e/(2q_T)}. \tag{35}$$

Therefore, under this approximation, the average output charge $\langle q\rangle$ can be determined by finding the positive root of the following equation:

$$\alpha_{\rm p}\gamma_0\, x^2 + (\alpha_{\rm n} - \alpha_{\rm p})\, x - \alpha_{\rm n}\gamma_0 = 0, \tag{36}$$

where $\gamma_0 = \exp(-\Delta V/(2V_T))$, and we have defined $x = \exp(\langle q\rangle_{\rm st}/q_T)$. This is precisely the same solution one finds with a classical deterministic analysis of the circuit (i.e., by finding the *fixed* value of the output charge or voltage that makes the currents through both transistors equal). Thus, we see that a deterministic analysis is only compatible with the assumption that the fluctuations in the circuit are the same as in equilibrium. To quantify the deviations of the actual fluctuations from equilibrium we define quantities $a$ and $b$ as follows:

$$\left\langle e^{\pm(q-\langle q\rangle_{\rm st})/q_T} \right\rangle_{\rm st} = e^{q_e/(2q_T)}e^{a\pm b}. \tag{37}$$

In this way, the parameters $a$ and $b$ account for the deviations with respect to equilibrium of the even central moments and odd central moments, respectively (if the stationary distribution is always symmetric around the mean value then $b = 0$). Then, given $a$ and $b$, the output charge $\langle q\rangle$ is determined by the positive root of:

$$\alpha_{\rm p}\gamma_0 e^{a+b}\, x^2 + (\alpha_{\rm n} - \alpha_p)\, x - \alpha_{\rm n}\gamma_0 e^{a-b} = 0. \tag{38}$$

## IV. FULL COUNTING STATISTICS OF THE CURRENTS

In this section we review a semi-analytical method to evaluate the characteristic function of the current fluctuations [4, 5]. The main quantities we will consider are the numbers $\{N_\rho(t)\}$ indicating the number of charges that went through a given device $\rho$ during a time $t$:

$$N_\rho(t) = \int_0^t d\tau I_\rho(\tau)/q_e, \tag{39}$$

where $I_\rho(\tau) = -q_e \sum_{\boldsymbol{q}} D_{\boldsymbol{q}}^\rho[\boldsymbol{j}]|_\tau$ is the stochastic electric current during a trajectory (see Section VI in the main text). The caracteristic function for these stochastic quantities is:

$$\Phi(\boldsymbol{\chi}) = \left\langle e^{i\sum_{\rho>0}\chi_\rho N_\rho} \right\rangle_{\boldsymbol{\mathcal{Q}}_t}, \tag{40}$$

where $\boldsymbol{\chi}$ is a vector of 'counting fields' $\chi_\rho > 0$, one for each device, and the average is taken over all trajectories compatible with a given initial distribution. The function $\Phi(\boldsymbol{\chi})$ can be evaluated in terms of a modified version of the generator associated to the master equation for the circuit. Before proceeding it is convenient to define this generator as follows. First, we write the master equation (Eq. (12) in the main text) in matrix form:

$$d_t|P(t)\rangle = -\hat{L}(t)|P(t)\rangle \qquad \text{with} \qquad |P(t)\rangle = \sum_{\boldsymbol{q}} P(\boldsymbol{q},t)|\boldsymbol{q}\rangle. \tag{41}$$

Thus, $\{|\boldsymbol{q}\rangle\}$ is a ortonormal basis in which each vector is associated to one particular circuit state ($\langle \boldsymbol{q}|\boldsymbol{q}'\rangle = \delta_{\boldsymbol{q},\boldsymbol{q}'}$), and $|P(t)\rangle$ is a vector encoding the probability distribution at time $t$. The generator $\hat{L}(t)$ is split into diagonal and non-diagonal parts:

$$\hat{L}(t) = \hat{\gamma}(t) - \hat{\Gamma}(t) \qquad \text{with} \qquad \hat{\gamma}(t) = \sum_{\boldsymbol{q}}|\boldsymbol{q}\rangle\langle\boldsymbol{q}|\,\gamma(\boldsymbol{q},t), \qquad \text{and} \qquad \hat{\Gamma}(t) = \sum_{\boldsymbol{q}}\sum_\rho |\boldsymbol{q}\rangle\langle\boldsymbol{q}-q_e\boldsymbol{\Delta}_\rho|\,\lambda_\rho(\boldsymbol{q}-q_e\boldsymbol{\Delta}_\rho). \tag{42}$$

6Then, the diagonal part $\hat{\gamma}(t)$ encodes the escape rates $\gamma(\bm{q},t) = \sum_\rho \lambda_\rho(\bm{q},t)$, while the non-diagonal part encodes the transition rates. Since the evolution given by $\hat{L}(t)$ preserves the normalization of the distribution $|P(t)\rangle$, it always has a zero eigenvalue, with a left eigenvector $|l_0\rangle = (1,1,\cdots,1)^T$ and a right eigenvector $|P_{\rm st}\rangle$, where $P_{\rm st}(\bm{q},t)$ is the stationary state, that we assume to be unique. Also, we will consider the modified generator:

$$L_{\bm{\chi}}(t) = \hat{\gamma}(t) - \hat{\Gamma}_{\bm{\chi}}(t), \qquad \text{with} \qquad \hat{\Gamma}_{\bm{\chi}}(t) = \sum_{\bm{q}} \sum_\rho |\bm{q}\rangle\langle \bm{q} - q_e \bm{\Delta}_\rho|\; \lambda_\rho(\bm{q} - q_e \bm{\Delta}_\rho)\, e^{i\chi_{|\rho|} s(\rho)}, \tag{43}$$

where $\chi_{|\rho|}$ is the counting field associated with device $\rho$ (note that the absolute value is necessary since in the previous expression the sum is over transitions, and therefore $\rho$ takes positive and negative values), and $s(\rho)$ is just the sign of $\rho$. It is important to note that the modified generator $\lambda_{\bm{\chi}}(t)$ does not preserve normalization. In fact, as shown in [4], the characteristic function defined in Eq. 40 can be expressed as:

$$\Phi(\bm{\chi}) = \langle l_0 | T_\tau e^{-\int_0^t d\tau L_{\bm{\chi}}(\tau)} | P(0) \rangle, \tag{44}$$

where $T_\tau$ is the time-ordering operator and $|P(0)\rangle$ is the initial state. For $\bm{\chi} = 0$, $U(t,0) = \exp(-\int_0^t d\tau L_{\bm{\chi}}(\tau))$ is just the evolution operator corresponding to the master equation. In that case, the previous equation is the sum of the probabilities in $|P(t)\rangle$ and therefore we have $\Phi(0) = 1$. For circuits with constant parameters we have the simpler result:

$$\Phi(\bm{\chi}) = \langle l_0 | e^{-t L_{\bm{\chi}}} | P(0) \rangle. \tag{45}$$

The previous expression can be numerically evaluated by truncating the state space to an appropiate number of dimensions and constructing the operator $L_{\bm{\chi}}$ in that truncated space. Once the characteristic function $\Phi(\bm{\chi})$ has been obtained, the probability distribution for the quantities $\{N_\rho(t)\}$ can be computed as:

$$P(\{N_\rho\}) = \int_{-\pi}^{\pi} \prod_{\rho>0} \frac{d\chi_\rho}{2\pi}\; \Phi(\bm{\chi})\, e^{-i \sum_{\rho>0} \chi_\rho N_\rho}. \tag{46}$$

This is how the results of Section VII A in the main text were obtained.

Finally, we mention that from the previous definitions it is possible to obtain a fluctuation theorem that is a special case of Eq. (18) in the main text. Indeed, for time-independent circuits and isothermal settings ($\beta_\rho = \beta$ for all $\rho$), we have the following symmetry of the operator $\hat{\Gamma}_{\bm{\chi}}$:

$$\hat{\Gamma}_{\bm{\chi}} = \hat{P}_{\rm eq}\, \hat{\Gamma}_{\overline{\bm{\chi}}}^T\, \hat{P}_{\rm eq}^{-1}, \tag{47}$$

where $\hat{P}_{\rm eq}$ is a diagonal operator encoding the equilibrium distribution $P_{\rm eq}(\bm{q}) = Z^{-1}\exp(-\beta\Psi(\bm{q}))$ (Eq. (79) in the main text), and $\overline{\bm{\chi}} = \{-\chi_\rho + iq_e\beta\,\Delta V_{n_f(\rho)}\}_{\rho>0}$, where $\Delta V_{n_f(\rho)}$ is the voltage difference associated to device $\rho$, as defined in Section V E in the main text. From this symmetry it follows that $\Phi(\bm{\chi}) = \Phi(\overline{\bm{\chi}})$, and consequently

$$\frac{P(\{N_\rho\})}{P(\{-N_\rho\})} = e^{q_e\beta \sum_{\rho>0} \Delta V_{n_f(\rho)} N_\rho}, \tag{48}$$

which is our final result.

## V. TRANSITION RATES FOR MOS TRANSISTORS WITH GROUNDED BODY

In this section we discuss the Poisson rates that one should assign to the drain-source conduction channel of a MOS transistor in which the body is grounded, instead of being connected to the source, as in Figure 3-(a). As discussed in the main text (recall Eq. (42)), in that case the symmetry between drain and source is preserved, but the mean current is not a function of the voltage bias $V_D - V_S$ only:

$$\langle I_D \rangle = I_0\, e^{(V_G - V_{\rm th})/(nV_T)}(e^{-V_S/V_T} - e^{-V_D/V_T}). \tag{49}$$

This is natural since the body potential introduces an internal voltage reference. However, the Poisson rates can still be considered functions of the voltage bias $V_D - V_S$ if we regard $V_D$ and $V_S$ as extra control parameters, in addition to $V_G$. Explicitly, we can consider the rates:

$$\begin{aligned}
\lambda_+ &= (I_0/q_e)\, e^{-V_{\rm th}/(nV_T)}\, e^{V_G/(nV_T)}\, e^{-V_D/V_T}\, e^{(V_D - V_S)/V_{\rm th}} \\
\lambda_- &= (I_0/q_e)\, e^{-V_{\rm th}/(nV_T)}\, e^{V_G/(nV_T)}\, e^{-V_S/V_T}\, e^{-(V_D - V_S)/V_{\rm th}}.
\end{aligned} \tag{50}$$

6


\thispagestyle{empty}



\end{document}